\documentclass[12pt]{article}

\usepackage{amsmath}
\usepackage{amssymb}
\usepackage[dvips]{graphicx}
\usepackage{setspace}

\newcommand{\ud}{\mathrm{d}}
\newcommand{\tsnr}{{\text{\footnotesize{SNR }}}}
\newcommand{\ssnr}{{\text{\scriptsize{SNR}}}}
\newcommand{\av}{{av}}
\newcommand{\K}{{\sf{K}}}
\newcommand{\C}{{\sf{C}}}
\newcommand{\I}{{\bf{I}}}
\newcommand{\vv}{{\bf{v}}}
\newcommand{\e}{\epsilon}

\newtheorem{prop:optphase}{Proposition}
\newtheorem{prop:Kuhn-Tucker}[prop:optphase]{Proposition}
\newtheorem{prop:discrete}[prop:optphase]{Proposition}
\newtheorem{prop:QAGC}[prop:optphase]{Proposition}
\newtheorem{prop:Rician}[prop:optphase]{Proposition}
\newtheorem{prop:Rician2}[prop:optphase]{Proposition}
\newtheorem{prop:twomass}[prop:optphase]{Proposition}
\newtheorem{prop:derivatives}[prop:optphase]{Proposition}
\newtheorem{cor:ebnozero}{Corollary}
\newtheorem{cor:widebandslope}[cor:ebnozero]{Corollary}
\newtheorem{conj:klessthenbeta}{Conjecture}
\newtheorem{prop:peakpower}[prop:optphase]{Proposition}
\newtheorem{prop:Ricianwithpeaklevel}[prop:optphase]{Proposition}
\newtheorem{prop:secondorderoptimal}[prop:optphase]{Proposition}
\newtheorem{def:BPSK}{Definition}
\newtheorem{def:QPSK}[def:BPSK]{Definition}
\newtheorem{prop:BPSK}[prop:optphase]{Proposition}
\newtheorem{prop:QPSK}[prop:optphase]{Proposition}
\newtheorem{cor:BPSKQPSKwidebandslope}[cor:ebnozero]{Corollary}
\newtheorem{prop:BPSKlimitedpeak}[prop:optphase]{Proposition}
\newtheorem{lemma:kurtosis}{Lemma}

\newtheorem{note:Rayleighfirstderiv}{Remark}
\newtheorem{note:derivativecomments}[note:Rayleighfirstderiv]{Remark}
\newtheorem{note:bitenergycomments}[note:Rayleighfirstderiv]{Remark}
\newtheorem{note:widebandslopecomments}[note:Rayleighfirstderiv]{Remark}
\newtheorem{note:limitedPARcomments}[note:Rayleighfirstderiv]{Remark}
\newtheorem{note:limitedpeakcomments}[note:Rayleighfirstderiv]{Remark}
\newtheorem{note:limitedpeakcomments2}[note:Rayleighfirstderiv]{Remark}
\newtheorem{note:limitedpeakcomments3}[note:Rayleighfirstderiv]{Remark}
\newtheorem{note:ooqpskcomments}[note:Rayleighfirstderiv]{Remark}
\newtheorem{note:ooqpskcomments2}[note:Rayleighfirstderiv]{Remark}
\newtheorem{note:limitedpeakoobpsk}[note:Rayleighfirstderiv]{Remark}

\begin{document}

\title{The Noncoherent Rician Fading Channel -- Part II :
Spectral Efficiency in the Low-Power Regime \footnote{This
research was supported by the U.S. Army Research Laboratory under
contract DAAD 19-01-2-0011. The material in this paper was
presented in part at the Fortieth Annual Allerton Conference on
Communication, Control, and Computing, Monticello, IL, Oct.,
2002.}}
\author{Mustafa Cenk Gursoy \and H. Vincent Poor \and Sergio Verd\'u}

\date{Dept. of Electrical
Engineering \\Princeton University \\[3.8pt] Princeton, NJ 08544}
\maketitle

\thispagestyle{empty}
\begin{spacing}{1.5}
%\begin{spacing}{1.1}
\begin{abstract}
Transmission of information over a discrete-time memoryless Rician
fading channel is considered where neither the receiver nor the
transmitter knows the fading coefficients. The
spectral-efficiency/bit-energy tradeoff in the low-power regime is
examined when the input has limited peakedness. It is shown that
if a fourth moment input constraint is imposed or the input
peak-to-average power ratio is limited, then in contrast to the
behavior observed in average power limited channels, the minimum
bit energy is not always achieved at zero spectral efficiency. The
low-power performance is also characterized when there is a fixed
peak limit that does not vary with the average power. A new
signaling scheme that overlays phase-shift keying on on-off keying
is proposed and shown to be optimally efficient in the low-power
regime.
\\

\end{abstract}
\end{spacing}

\begin{spacing}{1.5}
%\begin{spacing}{1.1}
\newpage
\section{Introduction} \label{sec:intro}
%\begin{spacing}{1.1}
\setcounter{page}{1} Emerging wireless systems, such as wideband
code division multiple access (WCDMA) and Impulse Radio, operate
at wide bandwidths. These systems can achieve higher data rates,
are more immune to the deleterious effects of multipath fading,
and require low power consumption. Many other wireless
communication systems such as satellite, deep space, and sensor
networks, also operate in the low-power regime where both spectral
efficiency (rate in bits per second divided by bandwidth in Hertz)
and energy-per-bit are low. For these systems, information
theoretic results on the spectral-efficiency/bit-energy tradeoff,
which reflects the fundamental tradeoff between bandwidth and
power, provide insightful results leading to the more efficient
use of resources in the low signal-to-noise ratio ($\tsnr\!\!$)
regime.

Verd\'u \cite{Verdu} has recently analyzed the spectral efficiency
of a general class of average power limited channels
characterizing the optimal bandwidth-power tradeoff in the
wideband regime. In particular, it is shown in \cite{Verdu} that
when the receiver has imperfect fading side information, input
signals with increasingly higher peak power is required to achieve
the capacity as $\tsnr \to 0$. On the other hand, limiting the
peakedness of the input signals, when neither the receiver nor the
transmitter knows the fading, is known to have a significant
impact on the achievable spectral efficiency in the low-power
regime \cite{Medard Gallager}, \cite{Telatar Tse},
\cite{Subramanian Hajek}. In this paper, we continue our study of
noncoherent Rician fading channels begun in Part I \cite{gursoy}
and consider the minimum energy per bit required for reliable
communication when the input signals have limited peakiness. The
organization of the paper is as follows. In Section
\ref{sec:prelim}, we review the basic measures of interest
in the low-power regime proposed in \cite{Verdu}.               %%%%%%%%%%%%%%%%%%%%%%%SV
In Section \ref{sec:specteff}, we analyze the
spectral-efficiency/bit-energy tradeoff in the noncoherent Rician
fading channel when, in addition to the average power limitation,
the input is subject to a fourth moment or a peak power
constraint. Finally, in Section \ref{sec:effsignal}, we show
efficient signaling schemes in the low-power regime, while Section
\ref{sec:conc} contains our conclusions.

\section{Preliminaries} \label{sec:prelim}
%%%%%%%%%%%%%%%%%%%%%%%%%%%%%%%%%%%%%%%%%%%%%%%%%%%%%%%%%%%%%%%%%%%
%%%%%%%%%%%%%%%  PRELIMINARIES  %%%%%%%%%%%%%%%%%%%%%%%%%%%%%%%%%%%
%%%%%%%%%%%%%%%%%%%%%%%%%%%%%%%%%%%%%%%%%%%%%%%%%%%%%%%%%%%%%%%%%%%

In the low-power regime, the spectral-efficiency/bit-energy
tradeoff is the key concept capturing the tradeoff between
bandwidth and power. We will denote the spectral efficiency as a
function of bit energy by ${\sf{C}}\left(\frac{E_b}{N_0}\right)$.
If we assume without loss of generality that one complex symbol
occupies a $\text{1s} \times \text{1Hz}$ time-frequency slot, then
the maximum achievable spectral efficiency can be obtained from
the Shannon capacity (bits/symbol)
\begin{gather}
{\sf{C}}\left(\frac{E_b}{N_0}\right) =
C(\textrm{\scriptsize{SNR}}) \quad \textrm{bits/s/Hz}
\end{gather}
where \vspace{-.5cm}
\begin{gather}
 \frac{E_b}{N_0} =
\frac{\textrm{\scriptsize{SNR}}}{C(\textrm{\scriptsize{SNR}})}
\end{gather}
is the bit energy normalized to the noise power. However, the
Shannon capacity, giving the full characterization of the
spectral-efficiency/bit-energy function, is either not known or
must be numerically computed for most fading channels whose
realizations are not fully known at the receiver. Hence one needs
to resort to approximation methods to examine the spectral
efficiency of a wide class of fading channels. In the low-\tsnr
regime, first-order linear approximation of the spectral
efficiency function provides an excellent match, and involves only
the slope of the spectral efficiency curve and the bit-energy at
zero spectral efficiency. The bit-energy at zero spectral
efficiency which depends only on the first derivative of the
capacity at zero {\footnotesize{SNR}}, i.e., \vspace{-.4cm}
\begin{gather}
\left.\frac{E_b}{N_0}\right|_{{\sf{C}} = 0} =
\lim_{\textrm{\scriptsize{SNR}} \rightarrow
0}\frac{\textrm{\scriptsize{SNR}}}{C(\textrm{\scriptsize{SNR}})} =
\frac{\log_e{2}}{\dot{C}(0)},
\end{gather}
is a relevant measure only in the asymptotic regime of infinite
bandwidth. Verd\'u \cite{Verdu} has recently given the following
formula for the wideband                                             %%%%%%%%%%%%%%%%%%SV
slope defined as the slope of the spectral efficiency curve
${\sf{C}}\left(\frac{E_b}{N_0}\right)$ in bits/s/Hz/3dB at zero
spectral efficiency:
\begin{eqnarray}
S_0 &\stackrel{\textrm{def}}{=}& \lim_{\frac{E_b}{N_0} \downarrow
\left.\frac{E_b}{N_0}\right|_{{\sf{C}} = 0}} \frac{{\sf{C}}\left(
\frac{E_b}{N_0}\right)}{10\log_{10}\frac{E_b}{N_0} - 10\log_{10}
\left. \frac{E_b}{N_0} \right|_{{\sf{C}} = 0}} \,\,10 \log_{10}2
\nonumber \\ &=&
\frac{2\left(\dot{C}(0)\right)^2}{-\ddot{C}(0)}\,\,,
\label{eq:widebanddef}
\end{eqnarray}
where $\dot{C}(0)$ and $\ddot{C}(0)$ denote the first and second
derivatives of capacity at zero {\footnotesize{SNR}} in nats. The
wideband slope closely approximates the growth of the spectral
efficiency curve in the low-power regime and hence is proposed as
a new tool providing insightful results when bandwidth is a
resource to be conserved.

For average power limited channels, the bit energy required for
reliable communications decreases  monotonically with decreasing
spectral efficiency, and the minimum bit energy is achieved at
zero spectral efficiency, $ \frac{E_b}{N_0}_{\text{min}} =
\left.\frac{E_b}{N_0}\right|_{{\sf{C}} = 0}$. Hence for fixed rate
transmission, reduction in the required power comes only at the
expense of increased bandwidth. By assuming an ergodic fading
process of finite second moment, Lapidoth and Shamai
\cite{Lapidoth Shamai} have shown that the minimum received bit
energy for an average power limited discrete-time single-input
single-output fading channel with Gaussian noise is $-1.59$ dB.
Recently, Verd\'u \cite{Verdu} has independently proven that the
minimum received bit energy of $-1.59$ dB is achieved in a general
class of average power limited multiple-input multiple-output
fading channels as long as the additive background noise is
Gaussian. This result holds regardless of the availability of the
fading knowledge at the receiver and transmitter. On the other
hand, it
is shown in \cite{Verdu} that having imperfect receiver side information has a  %%%%%%%%%%%%%%%SV
tremendous effect on the wideband slope. Although there is a
positive slope when there is perfect receiver channel side
information, imperfect fading knowledge at the receiver results in
zero wideband slope, and in this case flash signaling is required
to achieve the capacity in the low-power regime. Hence, achieving
the minimum bit energy becomes very demanding in terms of both
bandwidth and the peak-to-average ratio of the transmitted signal.

%%%%%%%%%%%%%%%%%%%%%%%%%%%%%%%%%%%%%%%%%%%%%%%%%%%%%%%%%%%%%%%%%%%
%%%%%%%%%%%%%%%  SPECTRAL EFFICIENCY AND BIT ENERGY  %%%%%%%%%%%%%%
%%%%%%%%%%%%%%%%%%%%%%%%%%%%%%%%%%%%%%%%%%%%%%%%%%%%%%%%%%%%%%%%%%%

\section{Spectral Efficiency vs. Bit Energy} \label{sec:specteff}

In this section, we consider the noncoherent Rician fading channel
model
\begin{gather} \label{eq:model}
y_i = mx_i + a_ix_i + n_i
\end{gather}
studied in Part I of this paper \cite{gursoy} and investigate the
spectral-efficiency versus energy-per-information bit tradeoff in
the low-power regime when the peakedness of the input signals is
limited by a fourth moment or a peak power constraint. Moreover,
at the end of this section, we comment on the spectral efficiency
in the low-\tsnr regime of the average power limited Rician
channel with phase noise.

\subsection{Second and Fourth Moment Limited Input} \label{subsec:2and4}

We first consider the case in which the input is subject to the
following second and fourth moment amplitude constraints:
\vspace{-0.05cm}
\begin{align}
E\{|x_i|^2\} &\leq P_{\av} \quad \forall i\label{eq:constraint1}
\\ E\{|x_i|^4\} &\leq \kappa P^2_{\av} \quad \forall i, \label{eq:constraint2}
\end{align}
where $1 < \kappa < \infty$. We have seen in the previous section
that the first-order linear approximation of the
spectral-efficiency/bit-energy function in the low-power regime is
determined by the first and second derivatives of the capacity at
zero {\footnotesize{SNR}}. Under quite general conditions on the
input and the channel, Prelov and Verd\'u \cite{Prelov Verdu}
obtained the exact asymptotic second-order behavior of the mutual
information between the channel input and the output for vanishing
{\footnotesize{SNR}}, \vspace{-0.2cm}
\begin{align} \label{eq:PrelovVerdu}
I(\bold{x},\bold{Hx+n}) = &E \left\{|| \bold{ \bar{H}}(\bold{x} -
E[\bold{x}] )  ||^2 \right\} \frac{\log e }{N_0} \nonumber + E
\left\{ \textrm{trace}\left (\textrm{cov}^2(\bold{Hx}|\bold{x})
\right)\right\}\frac{\log e}{2N_0^2} \nonumber \\ &-
\textrm{trace} \left(\textrm{cov}^2(\bold{Hx})\right)\frac{\log
e}{2N_0^2} + o(N_0^{-2})
\end{align}
where $\bold{H}$ is an $m \times n$ complex matrix of random
fading coefficients satisfying
$E\left\{||\mathbf{H}||^{4+\alpha}\right\} < \infty$ for some
$\alpha
> 0$, $\bar{\bold{H}} = E\{\bold{H}\}$, $N_0$ is the
one-sided noise spectral level, and $o(x)/x \to 0$ as $x \to 0$.
The only assumption on the input is that its probability
distribution satisfies \vspace{-.3cm}
\begin{gather}\label{eq:condontheinput}
P(\|\mathbf{x}\| > \delta) \leq \exp\{-\delta^v\}
\end{gather}

\vspace{-.3cm}\noindent for all sufficiently large $\delta > 0$,
where $v>0$ is a positive constant. Using the above result, we can
obtain the first and second derivatives of the mutual information
at zero {\footnotesize{SNR}} for the noncoherent Rician fading
channel (\ref{eq:model}) with second and fourth moment input
constraints.
%%%%%%%%%%%%%%%%%%%%%%%%%%%%%%%%%%%%%%%%%%%%%%%%%%%%Corollary derivatives of capacity
%%%%%%%%%%%%%%%%%%%%%%%%%%%%%%%%%%%%%%%%%%%%%%%%%%%%
%%%%%%%%%%%%%%%%%%%%%%%%%%%%%%%%%%%%%%%%%%%%%%%%%%%%
\begin{prop:derivatives} \label{cor:derivatives}
For the Rician channel model (\ref{eq:model}) with the Rician
factor $\K = \frac{|m|^2}{\gamma^2} > 0$  and input constraints
(\ref{eq:constraint1}) and (\ref{eq:constraint2}), the first and
second derivatives of capacity at $\textrm{\footnotesize{SNR}} =
0$ are
\begin{align}
\dot{C}(0) = |m|^2 \quad \text{and} \quad \ddot{C}(0) =
\kappa\gamma^4 - (|m|^2+\gamma^2)^2, \label{eq:secondderiv}
\end{align}
respectively.
\end{prop:derivatives}
\emph{Proof}: For the noncoherent Rician fading channel model, we
have established in \cite{gursoy} that the capacity-achieving
input distribution has finite support. Hence, the optimal input
distribution satisfies the condition (\ref{eq:condontheinput}).
Therefore, the proposition follows easily by specializing the
general result (\ref{eq:PrelovVerdu}) of Prelov and Verd\'u to the
Rician channel model (\ref{eq:model}), i.e.,
\begin{align}
I(x,mx+ax+n) = &|m|^2 \frac{E\{|x|^2\}}{N_0} + \frac{1}{2}
\gamma^4 \frac{E\{|x|^4\}}{N_0^2} -
\frac{1}{2}\left(|m|^2+\gamma^2\right)^2\left(\frac{
E\{|x|^2\}}{N_0}\right)^2 + o(N_0^{-2}).
\label{eq:secondorderasymp}
\end{align}
Since, by our assumption, $|m| > 0$, the first term on the right
hand side of (\ref{eq:secondorderasymp}) is maximized by having
$E\{|x|^2\} = P_{\av}$.  Note that since the other terms
characterize the second- and higher-order behavior, the average
power constraint should be satisfied with equality to achieve the
first derivative of the capacity at zero \tsnr. Notice also that
the second term is maximized by having $E\{|x|^4\} = \kappa
P_{\av}^2$, and hence the asymptotic capacity expression becomes
\begin{eqnarray}
C &=& |m|^2 \left(\frac{P_{\av}}{N_0}\right) + \frac{1}{2}
\left(\kappa \gamma^4 - \left(|m|^2+\gamma^2\right)^2\right)
\left(\frac{P_{\av}}{N_0}\right)^2 + o(N_0^{-2}) \nonumber
\\ &=& |m|^2\textrm{\scriptsize{SNR}} +
\frac{1}{2}\left(\kappa\gamma^4-\left(|m|^2+\gamma^2\right)^2\right)\textrm{\scriptsize{SNR}}^2
+ o(\textrm{\scriptsize{SNR}}^2), \label{eq:2nd_asympt}
\end{eqnarray}
from which the result follows. \hfill $\square$
%%%%%%%%%%%%%%%%%%%%%%%%%%%%%%%%%%%%%%%%%%%%%%%%%%%end Corollary
%%%%%%%%%%%%%%%%%%%%%%%%%%%%%%%%%%%%%%%%%%%%%%%%%%%%
%%%%%%%%%%%%%%%%%%%%%%%%%%%%%%%%%%%%%%%%%%%%%%%%%%%%
\begin{note:Rayleighfirstderiv}
\emph{It can be easily seen from the asymptotic expression
(\ref{eq:secondorderasymp}) that for the Rayleigh channel ($\K =
0$) with input constraints (\ref{eq:constraint1}) and
(\ref{eq:constraint2}), $\dot{C}(0) = 0$. We also note that Rao
and Hassibi \cite{RH03} have recently obtained the second-order
asymptotic expression of the mutual information in the multiple
antenna Rayleigh block fading channel when the fourth-order moment
of the input is finite, and shown that the mutual information is
zero to first order in \tsnr.}
\end{note:Rayleighfirstderiv}
\begin{note:derivativecomments}
\emph{Note that the first derivative depends only on the strength
of the line of sight component, $|m|^2$. Hence, for fixed $|m|$,
capacity curves under fourth moment constraints with different but
finite $\kappa$ values have the same slope at zero
{\footnotesize{SNR}}. On the other hand, the second derivative
depends on both $|m|^2$ and $\kappa$ and is positive if $\kappa >
(1+\K)^2$ where $\K = \frac{|m|^2}{\gamma^2}$ is the Rician
factor. Therefore, unlike the average power limited channels where
the capacity is a concave function of the {\footnotesize{SNR}},
the capacity curve in this case is a convex function locally
around $\textrm{\footnotesize{SNR}} = 0$. Finally, as a
comparison, if there is no fourth moment constraint (i.e., $\kappa
= \infty$), it is shown in \cite{Verdu} that $\dot{C}(0) = |m|^2 +
\gamma^2 \,\,,\,\, \ddot{C}(0) = -\infty.$}
\end{note:derivativecomments}
The above proposition reveals an interesting property. According
to \cite{Verdu_cost},
\begin{align}\label{eq:divergence}
\dot{C}(0) = |m|^2 &= N_0 \lim_{|x_0| \rightarrow 0}
\frac{D(f_{y|x=x_0}||f_{y|x = 0})}{|x_0|^2} \\ &= N_0 \lim_{|x_0|
\rightarrow 0} \frac{\frac{|m|^2 + \gamma^2}{N_0} |x_0|^2 -
\log_e\left(\frac{\gamma^2}{N_0}|x_0|^2+1 \right)}{|x_0|^2},
\end{align}
which shows that any on-off signaling scheme that satisfies the
second and fourth moment input constraints and whose on level
approaches the origin as $\textrm{\footnotesize{SNR}} \rightarrow
0$ achieves the first derivative of the capacity.

Moreover, imposing a fourth moment or a peak power constraint is
essentially the same as far as the first derivative of the
capacity is considered. In \cite{Prelov Meulen}, it is shown that
for a general class of memoryless channels with average and peak
power constraints, $ E\{|x|^2\} \leq P_{\av} \,\,,\,\, |x|^2
\stackrel{\text{a.s.}}{\leq} \kappa P_{\av}$ where $\kappa <
\infty$, the capacity has the following asymptotic expression as
$\text{\footnotesize{SNR}} \rightarrow 0$:
\begin{gather}
C(\tsnr) = \frac{1}{2} N_0 \, \Lambda \, \tsnr + o(\tsnr\!\!)
\end{gather}
where $\Lambda$ is the largest eigenvalue of the Fisher
information matrix \cite{Blahut}. We note that the limiting
expression on the right hand side of (\ref{eq:divergence}) is
equal to one half the largest eigenvalue of the Fisher information
matrix which, in our case, is $K = E_{f_{y|x = 0}}\{\vv(y,0)
\vv(y,0)^T\} = \frac{2|m|^2}{N_0} \, \I$ where
\begin{gather}
\vv(y,x) = \left[
\begin{array}{l}
\frac{\partial\log f_{y|x}}{\partial x_r} \\ \frac{\partial\log
f_{y|x}}{\partial x_i}
\end{array}
\right],
\end{gather}
$x_r$ and $x_i$ denote the real and imaginary parts of $x$
respectively, and $\I$ is the $2 \times 2$ identity matrix.
\par                                                                                %%%%%%%%%%%%%%%%%SV
For the average power limited Rician fading channel (without
fourth moment constraint) the following derivative can be
obtained:
$
\dot{C}(0) = |m|^2 + \gamma^2 = N_0 \lim_{|x_0| \rightarrow
\infty} \frac{D(f_{y|x=x_0}||f_{y|x = 0})}{|x_0|^2},
$
where in this case, the on level should escape to infinity to
achieve the first derivative of the capacity.

Having obtained analytical expressions for the first and second
derivatives of capacity at zero \textrm{\footnotesize{SNR}}, we
now find the bit energy required at zero spectral efficiency and
the wideband slope. We first note that the normalized received bit
energy in the Rician channel has the following formula:
\begin{gather}
\frac{E_b^r}{N_0} = \frac{E\{|m+a|^2\}\ssnr}{C(\ssnr)} =
\frac{(|m|^2+\gamma^2)\ssnr}{C(\ssnr)}.
\end{gather}

%%%%%%%%%%%%%%%%%%%%%%%%%%%%%%%%%%%%%%%%%%%%%%%%%%%%%%Bit energy at zero spectral efficiency Corollary
%%%%%%%%%%%%%%%%%%%%%%%%%%%%%%%%%%%%%%%%%%%%%%%%%%%%
%%%%%%%%%%%%%%%%%%%%%%%%%%%%%%%%%%%%%%%%%%%%%%%%%%%%
\begin{cor:ebnozero} \label{cor:ebnozero}
For the Rician fading channel (\ref{eq:model}) subject to input
constraints (\ref{eq:constraint1}) and (\ref{eq:constraint2}), the
normalized received bit energy $\frac{E_b^r}{N_0}$ required at
zero spectral efficiency and the wideband slope are
%\vspace{-0.2cm}
\begin{align}
\left. \frac{E_b^r}{N_0}\right|_{{\sf{C}} = 0} = \left(1 +
\frac{1}{\K}\right)\log_e2 \quad \text{and} \quad S_0 =
\frac{2\K^2}{\left(1+\K\right)^2-\kappa}, \label{eq:widebandslope}
\end{align}
respectively, where $\K = \frac{|m|^2}{\gamma^2}$ is the Rician
factor.
\end{cor:ebnozero}
\emph{Proof} : The received bit energy required at zero spectral
is obtained by letting $\tsnr \to 0$:
\begin{align}
\left. \frac{E_b^r}{N_0}\right|_{\C = 0} &= \lim_{\ssnr \to 0}
\frac{(|m|^2+\gamma^2) \ssnr}{C(\ssnr)} =
\frac{(|m|^2+\gamma^2)\log_e2}{\dot{C}(0)} =
\left(1+\frac{\gamma^2}{|m|^2}\right)\log_e2. \nonumber
\end{align}
The wideband slope expression is obtained by inserting the first
and second derivative expressions in (\ref{eq:secondderiv}) into
(\ref{eq:widebanddef}). Moreover, for the Rayleigh channel where
$\dot{C}(0) = 0$, it can be easily seen that
$\left.\frac{E_b^r}{N_0}\right|_{\C = 0} = \infty$ and $S_0 = 0$.
\hfill $\square$
%%%%%%%%%%%%%%%%%%%%%%%%%%%%%%%%%%%%%%%%%%%%%%end of corollary
%%%%%%%%%%%%%%%%%%%%%%%%%%%%%%%%%%%%%%%%%%%%%%%%%%%%
%%%%%%%%%%%%%%%%%%%%%%%%%%%%%%%%%%%%%%%%%%%%%%%%%%%%

\begin{note:bitenergycomments}
\emph{As long as a fourth moment constraint is imposed, the bit
energy required at zero spectral efficiency (or equivalently at
infinite bandwidth) depends only on the Rician factor $\K$, and
\begin{eqnarray}
\begin{array}{ll}
\left. \frac{E_b^r}{N_0}\right|_{{\sf{C}}= 0} \rightarrow \infty &
\textrm{as}  \quad {\sf{K}} \rightarrow 0
\\ \left. \frac{E_b^r}{N_0}\right|_{{\sf{C}} = 0} \rightarrow -1.59
\, \textrm{dB} & \textrm{as} \quad {\sf{K}} \rightarrow \infty
\end{array}
\end{eqnarray}
which also appeals intuitively because as ${\sf{K}}$ increases,
the channel becomes more Gaussian, and for the unfaded Gaussian
channel $ \left. \frac{E_b^r}{N_0}\right|_{{\sf{C}} = 0} = \log_e2
= -1.59\,\textrm{dB}$. For the Rayleigh fading channel, i.e., $|m|
= 0$, the bit energy required at zero spectral efficiency is
infinite. Therefore reliable communications is not possible at
this point. This is in stark contrast with the behavior observed
in average power limited channels where the bit energy required at
zero spectral efficiency is indeed the minimum one. On the other
hand, for the Rician fading channel where $|m|
> 0$, the required bit energy is finite.}
\end{note:bitenergycomments}
\begin{note:widebandslopecomments}
\emph{For average power limited channels, the wideband slope is
always nonnegative. In the noncoherent Rician fading channel
subject to second and fourth moment input limitations, we again
observe a markedly different behavior. From
(\ref{eq:widebandslope}), we see that if $\kappa >
\left(1+\K\right)^2$, then the wideband slope is negative, leading
to the conclusion that the minimum bit energy is achieved at a
nonzero spectral efficiency
${\sf{C}}^* > 0$. In this case, as observed in \cite[p.1341]{Verdu}, one should avoid operating in the  %%%%%%%%%%%%%SV
region where the spectral efficiency is lower than ${\sf{C}}^*$
because decreasing the spectral efficiency further in this region
(i.e., increasing the bandwidth for fixed rate transmission) only
increases the required power.}
\end{note:widebandslopecomments}
The following bounds on the minimum received bit energy are easily
obtained: $ \log_e2 \,\, \leq \,\, \frac{E_b^r}{N_0}_{\text{min}}
\leq \left(1+\frac{1}{\K}\right)\log_e2, $ where the lower bound
is the minimum received bit energy when there is only an average
power constraint, and the upper bound is the received bit energy
required at zero spectral
efficiency when the input is subject to a fourth moment constraint. %The
%bounds are uniform in the sense that they do not depend on the
%kurtosis constraint $\kappa$.
Note that the upper bound is loose in the Rayleigh case where $\K
= 0$. However, the larger the Rician factor $\K$, the smaller the
gap between the upper
and lower bounds. %For example, if ${\sf{K}} = 5$, then the upper
%bound is $-1.42 \, \textrm{dB}$.
If $\kappa \leq \left(1+\K\right)^2$, then the wideband slope is
positive and, based on numerical evidence, we conjecture that for
large enough value of $\K$, the minimum bit energy is achieved at
zero spectral efficiency and is equal to the bit energy expression
in (\ref{eq:widebandslope}).
%%%%%%%%%%%%%%%%%%%%%%%%%%%%%%%%%%%%%%%%%%%%%%%%%%%%CONJECTURE FOR K<(1+BETA^2)^2
%%%%%%%%%%%%%%%%%%%%%%%%%%%%%%%%%%%%%%%%%%%%%%%%%%%%
%%%%%%%%%%%%%%%%%%%%%%%%%%%%%%%%%%%%%%%%%%%%%%%%%%%%

\begin{figure}
\begin{center}
\includegraphics[width = 0.7\textwidth]{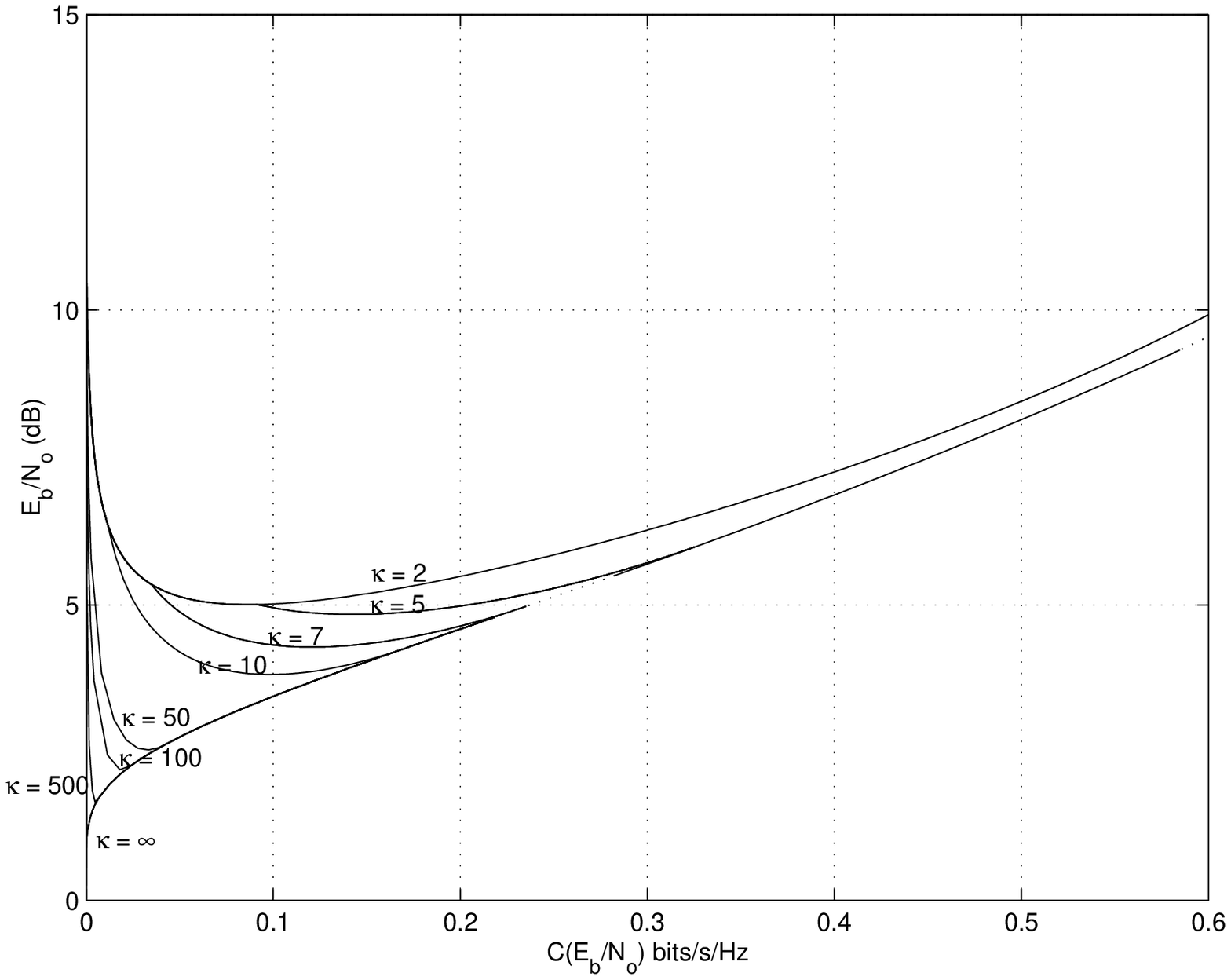}
\caption{$\frac{E_b}{N_0} \, (\text{\footnotesize{dB}}) $ vs.
Spectral Efficiency $C(\frac{E_b}{N_0})$ bits/s/Hz for the
Rayleigh Channel. ${\sf{K}} = 0$} \label{fig:specteffbeta0}
\end{center}
\end{figure}

\begin{figure}
\begin{center}
\includegraphics[width = 0.7\textwidth]{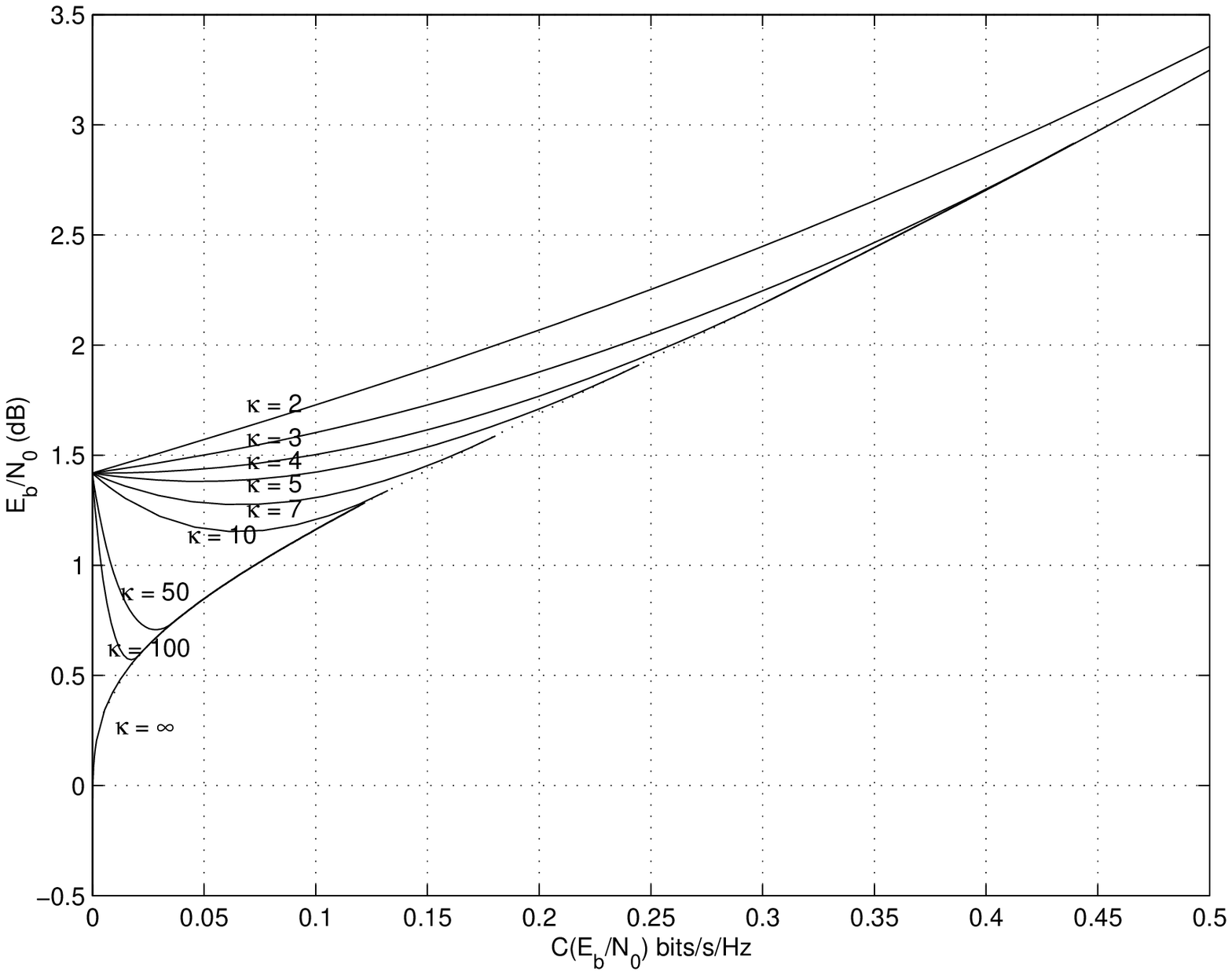}
\caption{$\frac{E_b}{N_0} \, (\textrm{\footnotesize{dB}}) $ vs.
Spectral Efficiency $C(\frac{E_b}{N_0})$ bits/s/Hz for the Rician
Channel with ${\sf{K}} = 1$.} \label{fig:specteffbeta1}
\end{center}
\end{figure}

\begin{figure}
\begin{center}
\includegraphics[width = 0.7\textwidth]{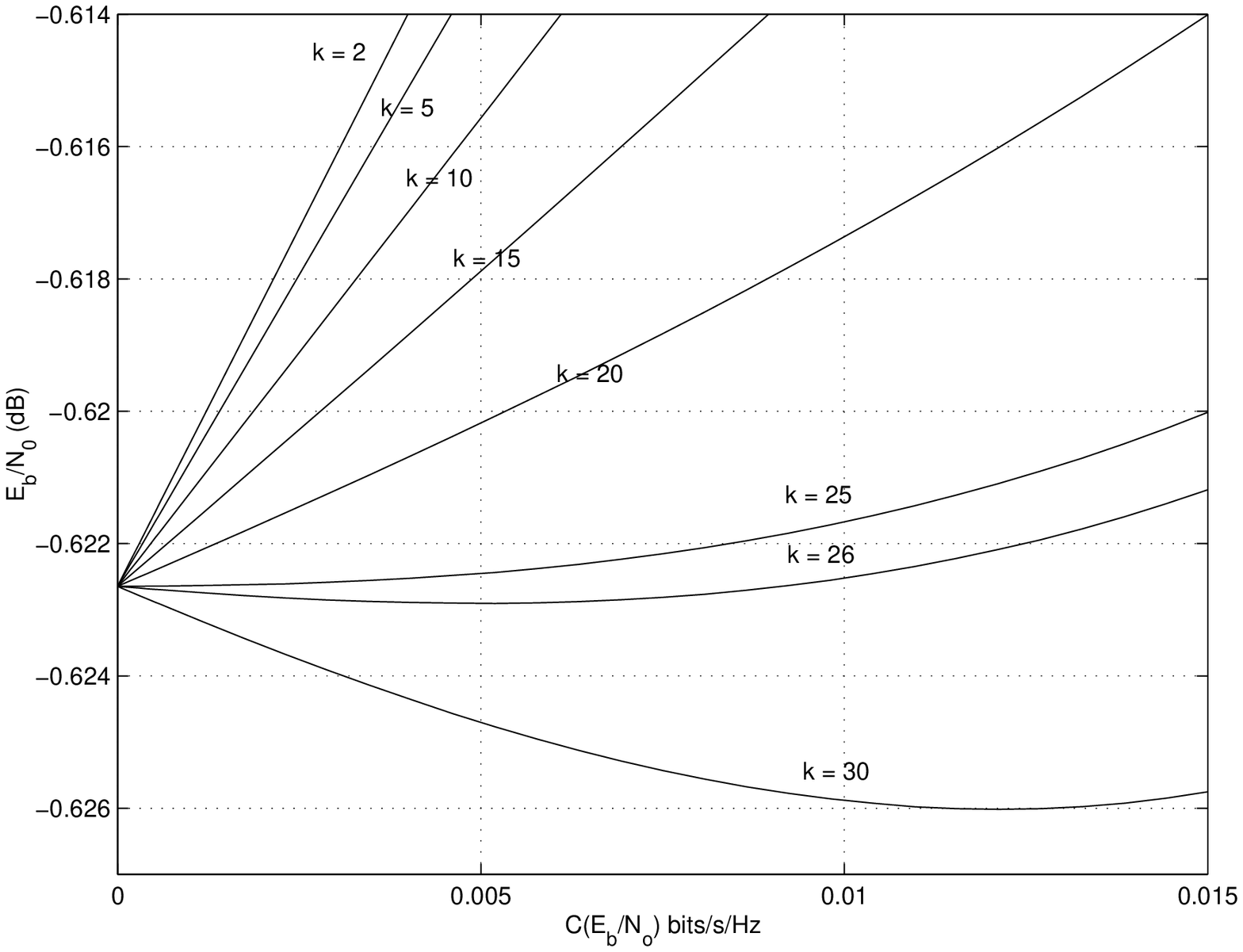}
\caption{$\frac{E_b}{N_0} \, (\textrm{\footnotesize{dB}}) $ vs.
Spectral Efficiency $C(\frac{E_b}{N_0})$ bits/s/Hz for the Rician
Channel with ${\sf{K}} = 2$.} \label{fig:specteffbeta2}
\end{center}
\end{figure}

Figures \ref{fig:specteffbeta0} and \ref{fig:specteffbeta1} plot
the $\frac{E_b^r}{N_0} \, (\textrm{\footnotesize{dB}}) $ vs.
${\sf{C}}(\frac{E_b^r}{N_0})$ bits/s/Hz curves for the Rayleigh
and Rician (${\sf{K}} = 1 $) channels, respectively, for various
values of $\kappa$. In the Rayleigh fading channel, for any finite
$\kappa$, the bit energy curve is bowl-shaped, achieving its
minimum at a nonzero spectral efficiency ${\sf{C}}^*$. Therefore,
for any $\frac{E_b}{N_0} > \frac{E_b}{N_0}_{\text{min}}$, there
are two spectral efficiencies ${\sf{C}}_1 < {\sf{C}}_2$ such that
$\frac{E_b}{N_0} = \frac{E_b}{N_0}({\sf{C}}_1) =
\frac{E_b}{N_0}({\sf{C}}_2)$. In this case, one should avoid the
low-power regime and operate at ${\sf{C_2}}$ where for the same
power and rate, less bandwidth is required. In the Rician fading
channel (${\sf{K}} = 1$), we observe the same behavior when
$\kappa > \left(1+\K\right)^2 = 4$. The minimum bit energy is
achieved at a nonzero spectral efficiency. However note that now
the bit energy required at zero spectral efficiency is finite and
is the same for all finite $\kappa$. If $\kappa \leq 4$, the bit
energy decreases monotonically with decreasing spectral efficiency
and the minimum bit energy is achieved at zero spectral
efficiency. Therefore in this case, the bandwidth-power tradeoff
is the usual one that we encounter in average power limited
channels; i.e., for fixed rate transmission, increasing the
bandwidth decreases the power required for reliable
communications. Similar conclusions are drawn by observing Fig.
\ref{fig:specteffbeta2} where the Rician factor has increased to
${\sf{K}} = 2$. We note that all the minimum bit energy points
other than that attained in the Rayleigh channel with $\kappa = 2$
are achieved by a two-mass-point distribution in the following
form: \vspace{-.4cm}
\begin{gather}
F(|x|) = \left(1-\frac{1}{\kappa}\right) u(|x|) + \frac{1}{\kappa}
u(|x| - \sqrt{\kappa N_0 \ssnr}). \label{eq:two-mass}
\end{gather}
An interesting observation is that in the Rayleigh channel for
sufficiently low $\tsnr$ values, the optimal input satisfies
$E\{|x|^2\} < P_\av$ and $E\{|x|^4\} = \kappa P_\av^2$, and hence
has a kurtosis higher than $\kappa$. In the case where $\kappa = 2
$, a two-mass-point distribution with $E\{|x|^2\} < P_\av$
achieves the minimum bit energy. Note that for the Rayleigh
channel, the second-order asymptotic term in
(\ref{eq:secondorderasymp}) is increased by decreasing the second
moment while satisfying the fourth moment constraint with equality

%%%%%%%%%%%%%%%%%%%%%%%%%%%%%%%%%%%%%%%%%%%%%%%%%%%%%%%%%%%%%%%%%%%%%%%%
%%%%%%%%%%%%%%%%%%%%%%%%%%%%%%%%%%%%%%%%%%%%%%%%%%%%%%%%%%%%%%%%%%%%%%%%
%%%%%%%%%%%%%%%%%%%%%%%%%%%%%%%%%%%%%%%%%%%%%%%%%%%%%%%%%%%%%%%%%%%%%%%%

\subsection{Average and Peak Power Limited Input}

In this section, we impose a peak power constraint, which is a
more stringent approach than constraining the fourth moment of the
amplitude. We analyze two cases: limited peak-to-average power
ratio and limited peak power. In contrast to the first case, no
constraint on the peak-to-average ratio is imposed in the second
case where the input is subject to a fixed peak power limit that
does not vary with the average power constraint.

\subsubsection{Limited Peak-to-Average Power Ratio}

We first consider the case in which the transmitter
peak-to-average power ratio is limited, and hence the input, in
addition to the average power constraint (\ref{eq:constraint1}),
is subject to %\vspace{-.5cm}
\begin{align}
%E\{|x_i|^2\} \leq P_{\av} \quad \forall i\label{eq:avgpower}\\
|x_i|^2 \stackrel{\text{a.s.}}{\leq} \kappa P_{\av} \quad \forall
i \label{eq:peakpower}
\end{align}
where $1 \leq \kappa < \infty$. The following result characterizes
the spectral-efficiency/bit-energy tradeoff in the low power
regime.
\begin{prop:peakpower}
For the Rician fading channel (\ref{eq:model}) with average and
peak power limitations (\ref{eq:constraint1}) and
(\ref{eq:peakpower}) respectively, the normalized received bit
energy $\frac{E_b^r}{N_0}$ required at zero spectral efficiency
and the wideband slope are
\begin{align}
\left. \frac{E_b^r}{N_0}\right|_{{\sf{C}} = 0} = \left(1 +
\frac{1}{\K}\right)\log_e2 \quad \text{and} \quad S_0 =
\frac{2\K^2}{\left(1+\K\right)^2-\kappa},
\label{eq:widebandslopelimitedPAR}
\end{align}
respectively, where $\K = \frac{|m|^2}{\gamma^2}$ is the Rician
factor.
\end{prop:peakpower}
\emph{Proof}: Note that since the input is subject to a peak
constraint (\ref{eq:peakpower}), the condition
(\ref{eq:condontheinput}) is satisfied, and hence the input-output
mutual information has again the same asymptotic expression
(\ref{eq:secondorderasymp}). It is easily observed from
(\ref{eq:secondorderasymp}) that in the Rayleigh channel where
$|m| = 0$, $\dot{C}(0) = 0$. If $\K
> 0 $, then to achieve the first derivative of the capacity at zero \tsnr,
we must have $E\{|x|^2\} = P_\av$. The following lemma gives the
maximum value of the fourth moment of the amplitude when the input
is subject to (\ref{eq:constraint1}) and (\ref{eq:peakpower}).
\begin{lemma:kurtosis} \label{lemma:kurtosis}
Consider a nonnegative real random variable $|x|$. Then we have
\begin{gather}\label{eq:maxfourthmoment}
\sup_{\substack{F_{|x|} \\ E\{|x|^2\} \leq P_{\av}
\\ |x|^2 \leq \kappa P_{\av} \,\,a.s.}}E\{|x|^4\} = \kappa
P_{\av}^2\,.
\end{gather}
Furthermore, the two-mass-point discrete distribution
(\ref{eq:maxfourthmomentdist}) achieves this supremum.
\begin{eqnarray}\label{eq:maxfourthmomentdist}
F_0^*(|x|) = \left( 1-\frac{1}{\kappa}\right)u(|x|) +
\frac{1}{\kappa}u(|x|-\sqrt{\kappa P_{\av}}).
\end{eqnarray}
\end{lemma:kurtosis}
\emph{Proof}: Following the approach in \cite{Abou} to find the
Kuhn-Tucker condition, it is easily established that a sufficient
and necessary condition for the distribution $F_0$ to achieve the
supremum in (\ref{eq:maxfourthmoment}) is that there exists
$\lambda \geq 0$ such that
\begin{eqnarray}
|x|^4 - \lambda |x|^2 &\leq& M - \lambda P_{\av} \qquad \forall
|x| \in [0,\sqrt{\kappa P_{\av}}] \\ &=& M - \lambda P_{\av}
\qquad \forall |x| \in E_0
\end{eqnarray}
where $E_0$ is the set of points of increase of $F_0$ and $M$ is
the supremum value. The two-mass-point distribution $F_0^*$
defined in (\ref{eq:maxfourthmomentdist}) satisfies these
constraints and achieves $M = \kappa P_{\av}^2$. \hfill $\square$

By the above Lemma and the fact that the average power constraint
has to be satisfied with equality to achieve the first derivative,
the asymptotic capacity expression for the Rician channel with $\K
> 0$ becomes
\begin{gather}\label{eq:ppasympcapacity}
C(\ssnr) = |m|^2\ssnr +
\frac{1}{2}\left(\kappa\gamma^4-\left(|m|^2+\gamma^2\right)^2\right)\ssnr\!^2
+ o(\ssnr\!^2).
\end{gather}
from which we see that $\dot{C}(0) = |m|^2$ and $\ddot{C}(0) =
\kappa \gamma^4 - \left(|m|^2+\gamma^2\right)^2$. Then the result
in (\ref{eq:widebandslopelimitedPAR}) is easily obtained from
these derivative expressions similarly as in the proof of
Corollary \ref{cor:ebnozero}. \hfill $\square$

\begin{note:limitedPARcomments}
\emph{Note that with average and peak power constraints, we obtain
the same bit energy and wideband slope expressions as in
(\ref{eq:widebandslope}) where the input is subject to $E\{|x|^2\}
\leq P_{\av}$ and $E\{|x|^4\} \leq \kappa P_{\av}^2$. Thus, we
conclude that imposing a fourth moment (\ref{eq:constraint2}) or a
peak constraint in the form (\ref{eq:peakpower}) has the same
effect in the low-power regime.}
\end{note:limitedPARcomments}

\subsubsection{Limited Peak Power} \label{sec:limitedpeak}
In this section, we assume that the transmitter is limited in peak
power and there is no constraint on the peak-to-average power
ratio. Hence, the input, in addition to the average power
constraint (\ref{eq:constraint1}), is subject to \vspace{-.4cm}
\begin{gather}
|x_i|^2 \stackrel{\text{a.s.}}{\le} \nu \quad \forall i
\label{eq:peakconstraintlimitedpeak}
\end{gather}
where $\nu$ is a fixed peak limit that does not vary with the
average power constraint $P_{\av}$. Notice in this case that as
$P_{\av} \downarrow 0$, the peak-to-average ratio increases
without bound. Recently, Sethuraman and Hajek \cite{Hajek}
analyzed the capacity per unit energy of Gaussian fading channels
with memory under similar average and peak power constraints.
Considering the memoryless Rician fading channel, we obtain the
following result on the minimum bit energy and wideband slope.

\begin{prop:Ricianwithpeaklevel} \label{prop:Ricianwithpeaklevel}
For the Rician fading channel (\ref{eq:model}) with input
constraints (\ref{eq:constraint1}) and
(\ref{eq:peakconstraintlimitedpeak}), and fixed noise density
$N_0$\footnote{$N_0$ is fixed so that $\ssnr$ varies only with
$P_{av}$, and the peak $\ssnr$ constraint is kept constant at
$\frac{\nu}{N_0}$. }, the minimum received bit energy and wideband
slope are
\begin{gather}
\frac{E_b^r}{N_0}_{\text{min}} = \frac{\log_e2}{1
-\frac{1}{\K+1}\frac{\log_e(1+\eta)}{\eta}} %\label{eq:ebnomin}
\quad
\text{and} \quad S_0 = \left\{
\begin{array}{ll} \frac{2 \left(\eta(\K + 1)- \log_e(1+\eta)\right)^2} { - 1 +
\frac{1}{1-\eta^2} \exp\left( \frac{2\K \eta^2}{1 - \eta^2}\right)
I_0\left(\frac{2\K \eta}{1-\eta^2}\right)} & \eta < 1
\\0 & \eta \ge 1
\end{array} \right., \label{eq:widebandslopelimitedpeak}
\end{gather}
respectively, where $\K = \frac{|m|^2}{\gamma^2}$ is the Rician
factor, $\eta = \frac{\gamma^2}{N_0}\nu$ is the normalized peak
power limit, and $I_0$ is the zeroth order modified Bessel
function of the first kind. Moreover, input signaling that
satisfies
\begin{gather}
E\{|x_{\ssnr}|^2\} = P_{av} = N_0 \tsnr, \quad |x_\ssnr|^2
\stackrel{a.s.}{\le} \nu, \label{eq:optcondition1}\\
\intertext{and} \lim_{\ssnr \to 0} \frac{E\left\{|x_\ssnr|^2
\,\,1\{|x_\ssnr|^2 > \nu - \epsilon \}\right\}}{E\{|x_\ssnr|^2\}}
= 1 \quad \forall \epsilon>0 \label{eq:optcondition2}
\end{gather}
is necessary to achieve the minimum bit energy, and hence the
wideband slope.
\end{prop:Ricianwithpeaklevel}
\textbf{Proof}: See Appendix \ref{app:proofofprop3}.

\begin{note:limitedpeakcomments}
\emph{The minimum bit energy decreases to $-1.59$ dB as we
approach the unfaded Gaussian channel, i.e., $\K \to \infty$; or
the peak constraint is relaxed, i.e., $\nu \to \infty$. We also
note that for the Rayleigh case where $\K = 0$, the minimum bit
energy expression can easily be obtained as a special case of the
capacity per unit energy result of \cite{Hajek}. The wideband
slope is zero for $\eta \ge 1$ as in the average power limited
case, and hence we conclude that achieving the minimum bit energy
is extremely demanding in bandwidth.}
\end{note:limitedpeakcomments}
\begin{note:limitedpeakcomments2}
\emph{It is also interesting to note that as $\eta \downarrow 0$
while keeping the fading variance $\gamma^2$ fixed (i.e., $\nu
\downarrow 0$ or $N_0 \uparrow \infty$), $\frac{E_b^r}{N_0}_{\min}
\to \left(1 + \frac{1}{\K}\right)\log_e2$ and $S_0 \to \frac{2
\K^2}{(1+\K)^2}$. We notice that the limiting values are the
expressions for the bit energy at zero spectral efficiency and
wideband slope with $\kappa = 0$ in the limited peak-to-average
power ratio case.}
\end{note:limitedpeakcomments2}

\begin{note:limitedpeakcomments3}
\emph{A class of input signals that satisfy the conditions
(\ref{eq:optcondition1}) and (\ref{eq:optcondition2}) are
identified to be first order optimal, thereby achieving the
minimum bit energy. Noting that these conditions are also
necessary to achieve the wideband slope, we show in Appendix
\ref{app:proofofprop3} that the two-mass-point amplitude
distribution
\begin{gather}
F(|x|) = \left( 1 - \frac{P_\av}{\nu} \right) u(|x|) +
\frac{P_\av}{\nu}\,u(|x| - \sqrt{v}), \label{eq:optimalinput}
\end{gather}
achieves both the minimum bit energy and the wideband slope.
Indeed, a recent independent analysis by Huang and Meyn
\cite{Meyn} has shown that the two-mass-point distribution with
one mass at the peak level and the other at the origin is
capacity-achieving for sufficiently low {\tsnr} values for a
general class of channels, including the Rician channel, with
fixed peak constraints.}
\end{note:limitedpeakcomments3}

\subsection{Average Power Limited Rician Channel with Phase Noise}
\begin{figure}
\begin{center}
\includegraphics[width = 0.7\textwidth]{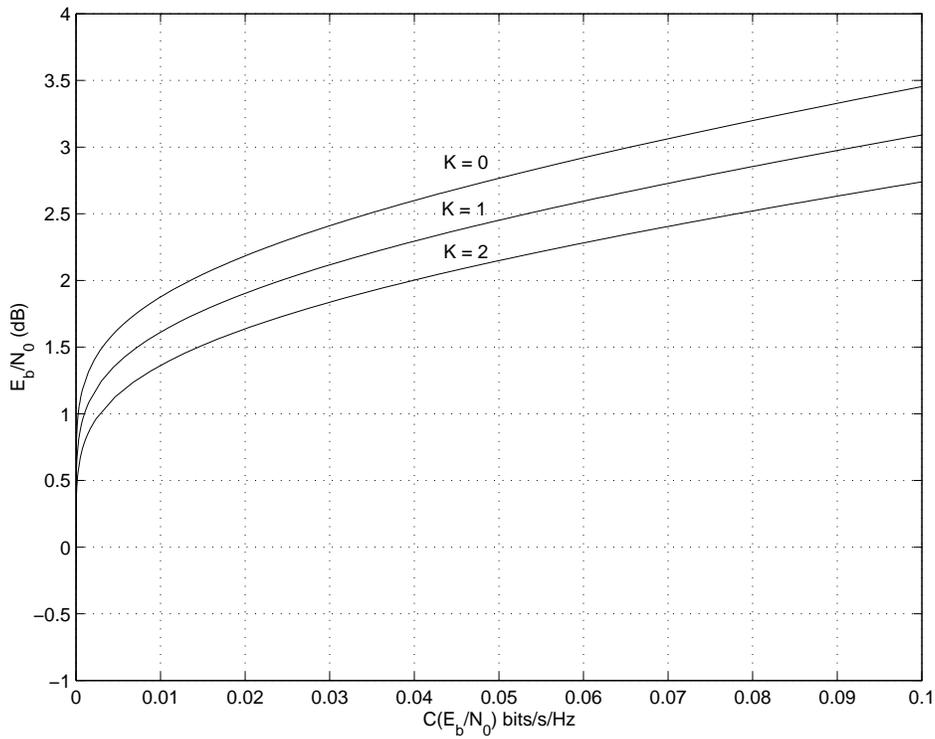}
\caption{$\frac{E_b}{N_0} \, (\textrm{\footnotesize{dB}}) $ vs.
Spectral Efficiency $C(\frac{E_b}{N_0})$ bits/s/Hz for Rician
fading channel with phase noise with ${\sf{K}} = 0,1,2$.}
\label{fig:pnK123}
\end{center}
\end{figure}

Finally, we comment on the spectral efficiency of the average
power limited Rician fading channel with phase noise which is
introduced in \cite{gursoy}. Lapidoth and Shamai \cite{Lapidoth
Shamai} have proven that for a general class of average power
limited single-input single-output fading channels, the bit energy
at zero spectral efficiency is $\log_e2 = -1.59 \,\text{dB}$. If
there is imperfect receiver side information, Verd\'u \cite{Verdu}
has shown that the wideband slope is zero. Figure \ref{fig:pnK123}
plots the spectral-efficiency/bit-energy function for the
noncoherent Rician channel with phase noise for ${\sf{K}} = 0,1$
and $2$. Indeed, we observe that the for all ${\sf{K}}$, the bit
energy curves are approaching $-1.59$ dB with zero slope.

%%%%%%%%%%%%%%%%%%%%%%%%%%%%%%%%%%%%%%%%%%%%%%%%%%%%%%%%%%%%%%%%%%%
%%%%%%%%%%%%%%%         EFFICIENT SIGNALING          %%%%%%%%%%%%%%
%%%%%%%%%%%%%%%%%%%%%%%%%%%%%%%%%%%%%%%%%%%%%%%%%%%%%%%%%%%%%%%%%%%

\section{Efficient Signaling in the Low-Power Regime} \label{sec:effsignal}

Having analyzed the spectral-efficiency/bit-energy tradeoff in the
low-power regime, we have seen that if the input is subject to
second and fourth moment constraints (\ref{eq:constraint1}) and
(\ref{eq:constraint2}), or average and peak power constraints
(\ref{eq:constraint1}) and (\ref{eq:peakpower}) with $\kappa \leq
\left(1+\K\right)^2$, then the wideband slope is positive and the
numerical results indicate that for large enough Rician factor
$\K$, the minimum bit energy is achieved at zero spectral
efficiency. Furthermore, if the noise spectral density $N_0$ and
the peak power constraint is fixed as the average power varies,
the capacity curve is a concave function of the \tsnr, and hence
the minimum bit energy is also achieved at zero spectral
efficiency. Motivated by these observations, we will now
investigate efficient signaling schemes in the low-power regime
when the input has limited peakedness. Verd\'u \cite{Verdu}
defines an input distribution to be \emph{first-order optimal} if
it satisfies the input constraints and achieves the first
derivative of the capacity at zero {\footnotesize{SNR}}, and
\emph{second-order optimal} if in addition it achieves the second
derivative of the capacity at zero {\footnotesize{SNR}}. So, a
first-order optimal input achieves the energy per bit at zero
spectral efficiency (which as noted before, need not be the
minimum energy per bit) and a second-order optimal input achieves
both the bit energy at zero spectral efficiency and the wideband
slope. We have observed in \cite{gursoy} that for the noncoherent
Rician fading channel with second and fourth moment input
constraints, a particular two-mass-point input distribution
(\ref{eq:two-mass}) is capacity-achieving for sufficiently small
{\footnotesize{SNR}}. Based on this observation, we define the
following signaling schemes which overlay phase-shift keying on on-off keying:           %%%%%%%%%%%%%%SV

\begin{def:BPSK}
An OOBPSK signal, parametrized by $0 < p \leq 1$, has the
following constellation points with the corresponding
probabilities \vspace{-0.1cm}
\begin{eqnarray} \label{eq:BPSK}
\begin{array}{lll} x_1 = 0 &\text{with prob.} &1-p \\ x_2 =
+\sqrt{P_{av}/p} &\qquad\!\!''& p/2 \\ x_3 = -\sqrt{P_{av}/p}
&\qquad\!\!''& p/2
\end{array}
\end{eqnarray}
where $P_{av}$ is the average power of the signal.
\end{def:BPSK}

\begin{def:QPSK}
An OOQPSK signal, parametrized by $0 < p \leq 1$, has the
following constellation points with the corresponding
probabilities
\begin{eqnarray} \label{eq:QPSK}
\begin{array}{lll} x_1 = 0 &\text{with prob.} &1-p \\ x_i =
\sqrt{\frac{P_{av}}{2p}}(\pm 1 \pm j) &\qquad\!\!'' &p/4 \qquad i
= 2,3,4,5
\end{array}
\end{eqnarray}
where $P_{av}$ is the average power of the signal.
\end{def:QPSK}

From the above definitions, we immediately notice that $1/p$ is
the kurtosis of the signals and having $p = 1$ reduces the
signaling schemes to ordinary BPSK and QPSK respectively. Next we
investigate the performance of these schemes in the wideband
regime when the input amplitude is subject to second and fourth
moment limitations.
%%%%%%%%%%%%%%%%%%%%%%%%%%%%%%%%%%%%%%%%%%%%%%%%%%%%OPTIMALITY OF GENERALIZED BPSK AND QPSK
%%%%%%%%%%%%%%%%%%%%%%%%%%%%%%%%%%%%%%%%%%%%%%%%%%%%
%%%%%%%%%%%%%%%%%%%%%%%%%%%%%%%%%%%%%%%%%%%%%%%%%%%%
\begin{prop:BPSK} \label{prop:BPSK}
For the Rician fading channel (\ref{eq:model}) with input
constraints (\ref{eq:constraint1}) and (\ref{eq:constraint2}), an
OOBPSK input with average power $P_{av}$ and $\frac{1}{\kappa}
\leq p \leq 1$ is first-order optimal. Furthermore the first and
second derivatives at zero {\footnotesize{SNR}} of the mutual
information achieved by this input are given by \vspace{-0.2cm}
\begin{align}
\dot{I}(0)= |m|^2 \quad \text{and} \quad \ddot{I}(0) =
\kappa\gamma^4 - (|m|^2+\gamma^2)^2 - |m|^4, \label{eq:BPSK2deriv}
\end{align}
respectively.
\end{prop:BPSK}
\emph{Proof}: First note that an OOBPSK input (\ref{eq:BPSK}) with
average power $P_{av}$ and $\frac{1}{\kappa} \leq p \leq 1$
satisfies the input constraints (\ref{eq:constraint1}) and
(\ref{eq:constraint2}). The input-output mutual information is
given by
\begin{gather} \label{eq:mutualinfo}
I(x,y) = \int_{\mathbb{C}} \int_{\mathbb{C}} f_{y|x}(y|x)
\ln\frac{f_{y|x}(y|x)}{f_y(y)} \, \ud y\, \ud F(x)
\end{gather}
where $ \ud F(x) = \left(1-p\right)\delta(x) + \frac{p}{2}
\,\delta\!\!\left(x - \sqrt{\frac{P_{av}}{p}}\right) +
\frac{p}{2}\,\delta\!\!\left(x + \sqrt{\frac{P_{av}}{p}}\right) $,
and $f_{y|x}$ is given in \cite[Eqn. 5]{gursoy}. Direct
differentiation of (\ref{eq:mutualinfo}) with respect to
{\footnotesize{SNR}} gives (\ref{eq:BPSK2deriv}). Achieving the
first derivative of capacity at zero {\footnotesize{SNR}}, OOBPSK
input is first order optimal. \hfill $\square$

\begin{prop:QPSK} \label{prop:QPSK}
For the Rician fading channel model (\ref{eq:model}) with input
constraints (\ref{eq:constraint1}) and (\ref{eq:constraint2}), the
OOQPSK input with average power $P_{av}$ and $p =
\frac{1}{\kappa}$ is second-order optimal, i.e., the first and
second derivatives at zero {\footnotesize{SNR}} of the mutual
information achieved by this input are given by \vspace{-0.3cm}
\begin{gather}
\dot{I}(0) = |m|^2 \quad \text{and} \quad \ddot{I}(0) =
\kappa\gamma^4 - (|m|^2+\gamma^2)^2, \label{eq:QPSK2deriv}
\end{gather}
respectively.
\end{prop:QPSK}
\emph{Proof}: The steps in the proof are essentially the same as
in the proof of Proposition \ref{prop:BPSK}. The OOQPSK input
(\ref{eq:QPSK}) with average power $P_{av}$ and $p=
\frac{1}{\kappa}$ satisfies the input constraints
(\ref{eq:constraint1}) and (\ref{eq:constraint2}), and the input
distribution in the mutual information expression
(\ref{eq:mutualinfo}) now becomes %\vspace{-0.3cm}
\begin{align}
\ud F(x) = &\left(1-p\right)\delta(x) +
\frac{p}{4}\,\delta\!\!\left(x -
\sqrt{\frac{P_{av}}{2p}}(1+j)\right) +
\frac{p}{4}\,\delta\!\!\left(x -
\sqrt{\frac{P_{av}}{2p}}(1-j)\right) \nonumber \\ &
\frac{p}{4}\,\delta\!\!\left(x -
\sqrt{\frac{P_{av}}{2p}}(-1+j)\right) +
\frac{p}{4}\,\delta\!\!\left(x -
\sqrt{\frac{P_{av}}{2p}}(-1-j)\right).
\end{align}
Similarly, direct differentiation of the mutual information with
respect to {\footnotesize{SNR}} provides (\ref{eq:QPSK2deriv}).
Achieving both the first and the second derivatives of capacity at
zero {\footnotesize{SNR}}, OOQPSK input is second-order optimal.
\hfill $\square$

\begin{note:ooqpskcomments}
\emph{Note that the results of Proposition \ref{prop:BPSK} and
\ref{prop:QPSK} hold when the fourth moment constraint
(\ref{eq:constraint2}) is replaced by a peak power constraint
(\ref{eq:peakpower}).}
\end{note:ooqpskcomments}

\begin{note:ooqpskcomments2}
\emph{An interesting observation is that OOBPSK with $p \in
[\frac{1}{\kappa}, 1)$ is, to first order, no better than ordinary
BPSK (i.e., OOBPSK with $p = 1$) because the first derivative does
not depend on the kurtosis of the signal. Since no information
transfer takes place with phase modulation over the unknown
Rayleigh fading channel, this explains why reliable communication
over this channel is not possible in the asymptotic regime of zero
spectral efficiency. For second order optimality, we need OOQPSK
signaling with $p = \frac{1}{\kappa}$. Therefore, in the low-power
regime, we need both amplitude and phase modulation schemes in
order to be spectrally efficient.}
\end{note:ooqpskcomments2}

For the average-power limited unfaded AWGN channel,
\cite{Verdu} has shown that ordinary QPSK, which has unit               %%%%%%%%%%%%%%%%%%%%SV
kurtosis, is second-order optimal. Hence, for the unfaded channel,
introducing an additional peak or fourth moment constraint on the
channel input does not degrade the performance in the wideband
regime.

As a natural next step, we look at the wideband slopes achieved by
OOBPSK and OOQPSK signaling.

\begin{cor:BPSKQPSKwidebandslope}
In the Rician fading channel (\ref{eq:model}) with input
constraints (\ref{eq:constraint1}) and (\ref{eq:constraint2}),
OOBPSK and OOQPSK signaling schemes with $p = \frac{1}{\kappa}$,
achieve the following wideband slopes :
\begin{align}
S_{0,\textrm{OOBPSK}} = \frac{2\K^2}{\K^2 +
\left(1+\K\right)^2-\kappa} \quad \text{and} \quad
S_{0,\textrm{OOQPSK}} = \frac{2\K^2}
{\left(1+\K\right)^2-\kappa}\,\,,
\end{align}
respectively.
\end{cor:BPSKQPSKwidebandslope}
Since the OOQPSK is second-order optimal, it achieves the optimal
wideband slope. Note also that, for fixed $\kappa$ and $\K$, when
both slopes are positive, OOBPSK achieves smaller slope than that
of OOQPSK. Hence for the same bit energy and rate, OOBPSK needs
more bandwidth in the low-power regime, or equivalently, for the
same bandwidth and rate it requires more bit energy. We need to be
careful when the slopes are negative because in this case we want
to avoid operating in the very low-power regime as discussed
previously.

\begin{figure}
\begin{center}
\includegraphics[width = 0.67\textwidth]{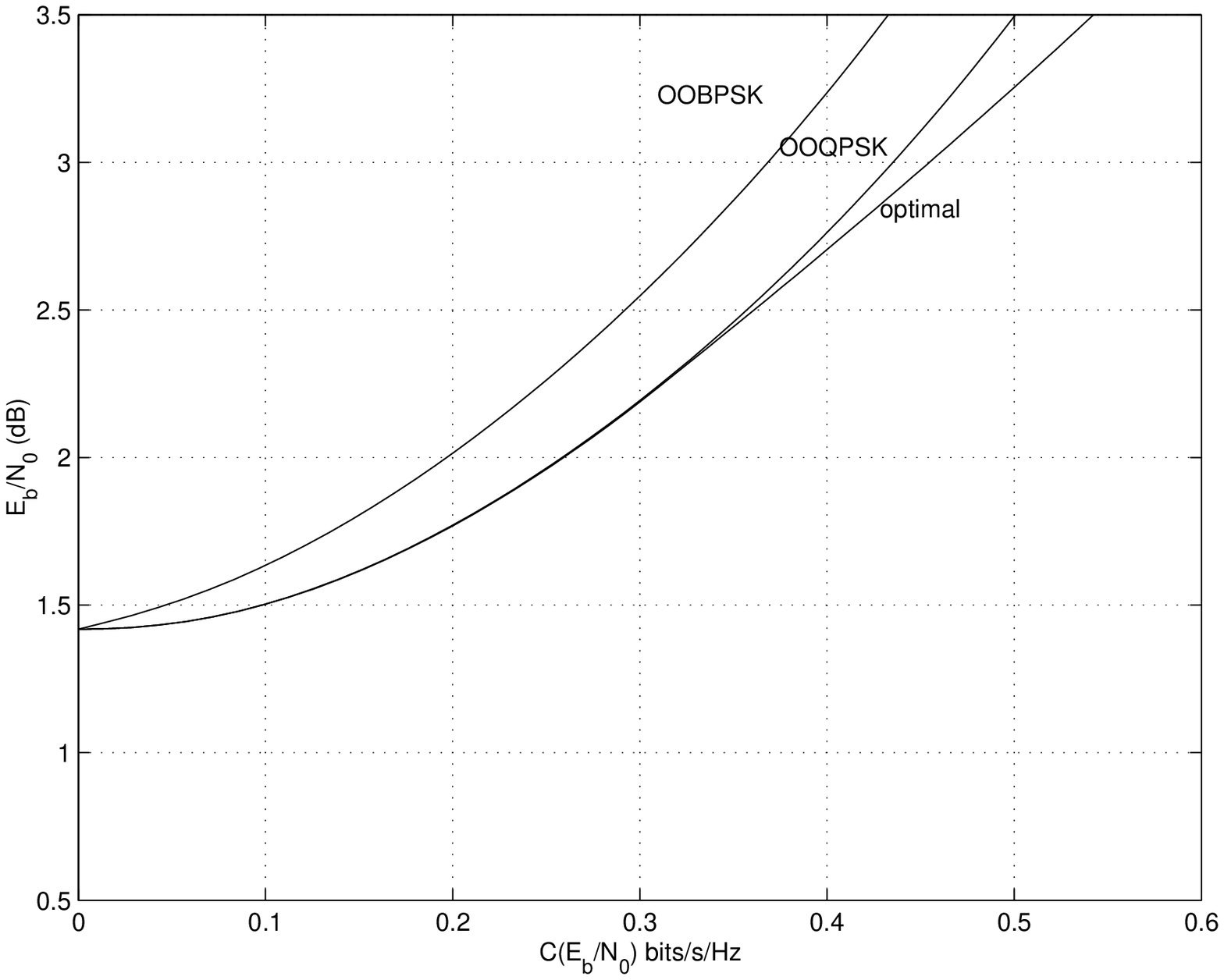}
\caption{$\frac{E_b}{N_0} \, (\textrm{\footnotesize{dB}}) $ vs.
Spectral Efficiency $C(\frac{E_b}{N_0})$ bits/s/Hz curves for the
optimal, OOQPSK and OOBPSK signalling with $p = \frac{1}{\kappa}$
in the Rician Channel where ${\sf{K}} = 1$ and $\kappa = 4$.}
\label{fig:obqbeta1k4}
\end{center}
\end{figure}

\begin{figure}
\begin{center}
\includegraphics[width = 0.7\textwidth]{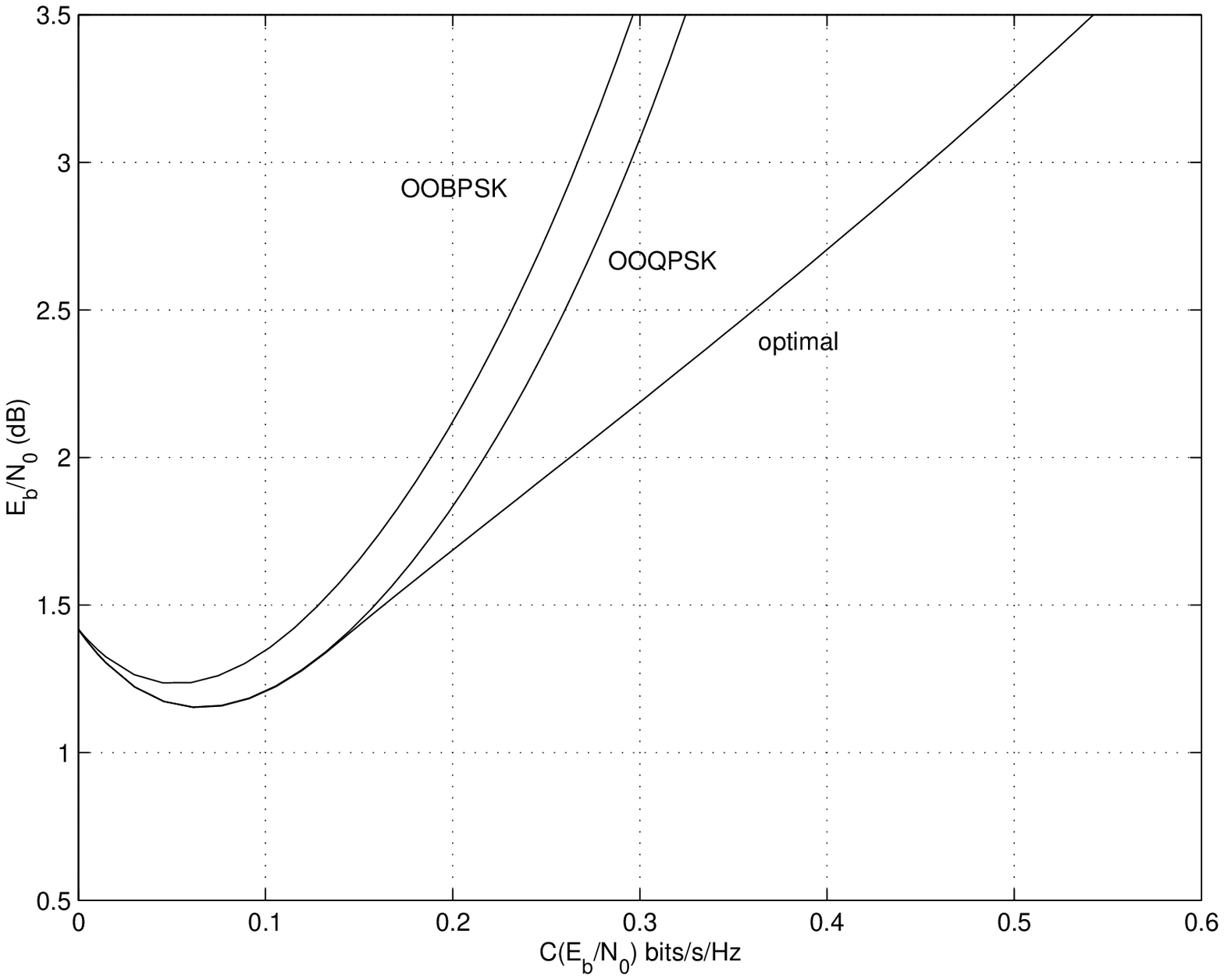}
\caption{$\frac{E_b}{N_0} \, (\textrm{\footnotesize{dB}}) $ vs.
Spectral Efficiency $C(\frac{E_b}{N_0})$ bits/s/Hz curves for
optimal, OOQPSK and OOBPSK signalling with $p = \frac{1}{\kappa}$
in the Rician Channel where ${\sf{K}} = 1$ and $\kappa = 10$.}
\label{fig:obqbeta1k10}
\end{center}
\end{figure}

\begin{figure}
\begin{center}
\includegraphics[width = 0.66\textwidth]{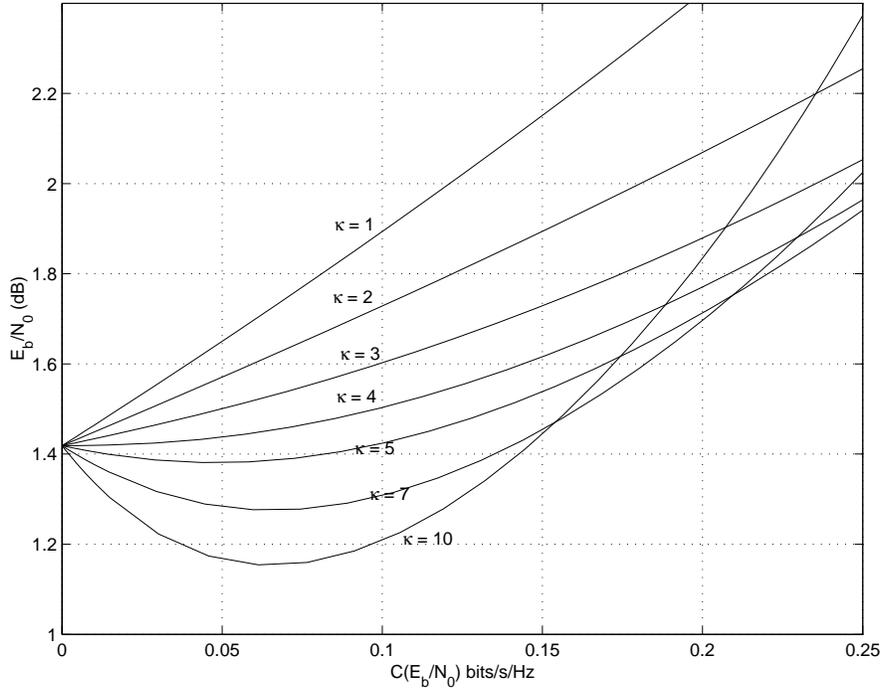}
\caption{$\frac{E_b}{N_0} \, (\textrm{\footnotesize{dB}}) $ vs.
Spectral Efficiency $C(\frac{E_b}{N_0})$ bits/s/Hz curves for
OOQPSK signalling for $\frac{1}{p}=\kappa = 1,2,3,4,5,7,10$ in the
Rician Channel with ${\sf{K}} = 1$.} \label{fig:qpskbeta1}
\end{center}
\end{figure}

\begin{figure}
\begin{center}
\includegraphics[width = 0.7\textwidth]{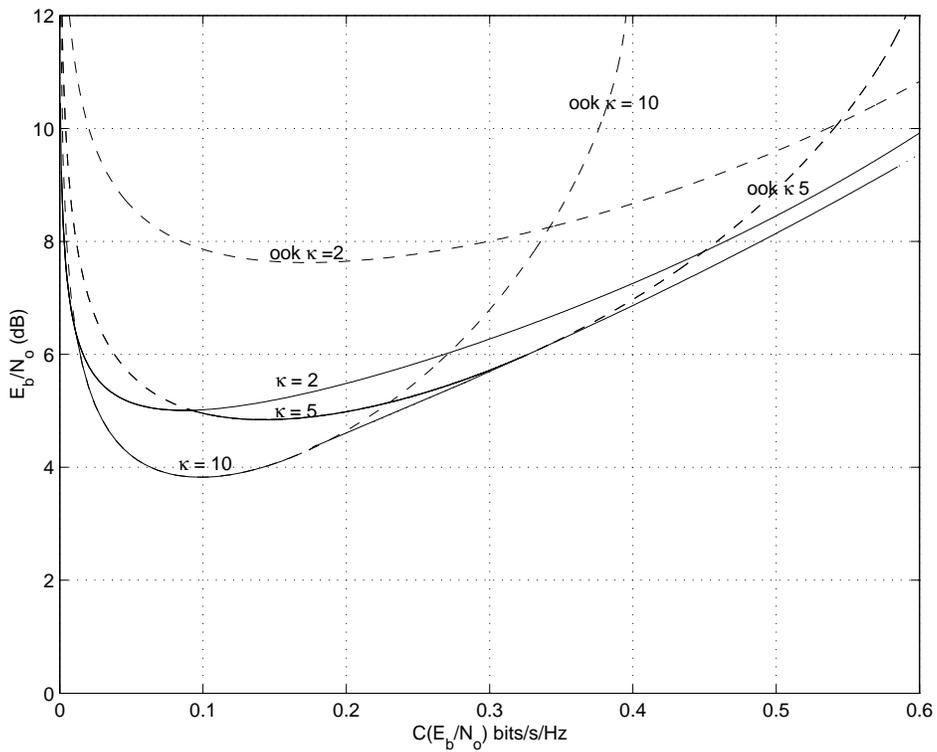}
\caption{Spectral Efficiency $C(\frac{E_b}{N_0})$ bits/s/Hz  vs.
$\frac{E_b}{N_0} \, (\textrm{\footnotesize{dB}})$ for the optimal
(solid curves) and the OOK signaling (dashed curves) with $\kappa
= 2,5,10$ in the Rayleigh Channel.} \label{fig:bpskbeta1}
\end{center}
\end{figure}

Figure \ref{fig:obqbeta1k4} plots the spectral efficiency-bit
energy curve for optimal, OOQPSK and OOBPSK signaling with $p =
\frac{1}{\kappa}$ in the Rician fading channel (${\sf{K}} = 1$)
when the input is subject to $E\{|x|^2\} \le P_\av$ and
$E\{|x|^4\} \le \kappa P_\av^2$ with $\kappa = 4$. Note that for
this value of $\kappa$, the wideband slope is positive. Both
OOBPSK and OOQPSK achieve the minimum bit energy (first order
optimality). Being second-order optimal, OOQPSK also achieves the
wideband slope and is very close to the optimal curve in the
low-\tsnr regime. Therefore, we conclude that OOQPSK signaling is
optimally efficient in the low-power limit. Note that OOBPSK
achieves a smaller slope and hence for fixed rate and power it
requires more bandwidth. In Fig. \ref{fig:obqbeta1k10}, $\kappa$
is increased to 10. In this case, the wideband slope is negative
and the minimum bit energy is achieved at a nonzero spectral
efficiency, which implies that the very low-power regime ought to
be avoided. However, we observe that OOQPSK is still an efficient
scheme achieving very close to the minimum bit energy. Moreover,
Fig. \ref{fig:qpskbeta1} shows how we increase the wideband slope
when we use OOQPSK signaling with higher and higher kurtosis
$\kappa$ as long as $\kappa \leq (1+\K)^2$. Finally for the
Rayleigh fading channel, Fig. \ref{fig:bpskbeta1} plots the
bit-energy/spectral-efficiency curve for the optimal and OOK
signaling, which is in the form given by (\ref{eq:two-mass}), for
$\kappa = 2,5$ and $10$. We observe that OOK is an efficient
signaling scheme achieving the minimum bit energy in the cases of
$\kappa = 5$ and $10$. For $\kappa = 2$, the minimum bit energy is
again achieved by on-off keying, however as discussed in Section
\ref{subsec:2and4}, with $E\{|x|^2\} < P_\av$. Therefore in this
case OOK signaling in the form (\ref{eq:two-mass}) is suboptimal
achieving a higher minimum bit energy.

Up to now, we have considered the cases in which the fourth moment
or the peak power constraint varies with the average power of the
signal. When there is a fixed peak limit, we have seen in Section
\ref{sec:limitedpeak} that the two-mass-point input amplitude
distribution (\ref{eq:optimalinput}) is required to achieve the
minimum bit energy. Therefore, when the input is subject to
(\ref{eq:constraint1}) and (\ref{eq:peakconstraintlimitedpeak}),
OOK signaling where the on-level is at $\sqrt{\nu}$ with
probability $\frac{P_\av}{\nu}$ is first-order optimal. Moreover,
it can be easily seen that this signaling scheme is second-order
optimal in the Rayleigh channel and in the Rician channel when the
wideband slope is zero i.e., $\nu \ge \frac{N_0}{\gamma^2}$. We
also investigate the low-power performance of OOQPSK signaling.
Note that by choosing $p = \frac{P_\av}{\nu}$, we obtain OOQPSK
signals whose on-level amplitudes are fixed at $\sqrt{\nu}$. Since
the proof of the following result is similar to those of
Proposition \ref{prop:BPSK} and \ref{prop:QPSK}, and only involves
differentiating the mutual information with respect to the \tsnr,
it is omitted. \vspace{-.2cm}
\begin{prop:BPSKlimitedpeak} \label{prop:BPSKlimitedpeak}
For the Rician fading channel (\ref{eq:model}) with input
constraints (\ref{eq:constraint1}),
(\ref{eq:peakconstraintlimitedpeak}) and fixed noise density
$N_0$, the OOQPSK input with average power $P_{av}$ and $p =
\frac{P_\av}{\nu}$ is first-order optimal. Furthermore the first
and second derivatives at zero {\footnotesize{SNR}} of the mutual
information achieved by this input are given by \vspace{-0.2cm}
\vspace{-.2cm}
\begin{gather*}
\dot{I}(0) = (|m|^2 + \gamma^2) - N_0 \frac{\log_e\left(
\frac{\gamma^2}{N_0}\nu + 1 \right)}{\nu}
\,\,\,\,\text{and}\,\,\,\, \ddot{I}(0) = \left \{
\begin{array}{ll}
\frac{N_0^2}{\nu^2}\left(1 - \frac{1}{16} \sum_{i,j = 2}^5 \int
\frac{f_{y|x = x_i} \, f_{y|x = x_j}}{f_{y|x = 0}} \, \ud y
\right) & \nu < \frac{N_0}{\gamma^2} \\ -\infty & \nu \ge
\frac{N_0}{\gamma^2}
\end{array}
\right.,
\end{gather*}
respectively, where $f_{y|x=a} = \frac{1}{\pi (\gamma^2 |a|^2 +
N_0)} \exp\left(-\frac{|y - ma |^2}{\gamma^2 |a|^2 + N_0}\right)$.
\end{prop:BPSKlimitedpeak}
\vspace{-.3cm}
\begin{note:limitedpeakoobpsk}
\emph{We note that the OOQPSK signaling is second-order optimal in
the Rician channel only when the wideband slope is zero i.e., $\nu
\ge \frac{N_0}{\gamma^2}$, and achieves a smaller slope if $\nu <
\frac{N_0}{\gamma^2}$}.
\end{note:limitedpeakoobpsk}

\section{Conclusion} \label{sec:conc}

For the noncoherent Rician fading channel, we have analyzed the
spectral-efficiency/bit-energy tradeoff in the low-power regime
when the input peakedness is limited by a fourth moment or a peak
power constraint. We have found analytical expressions for the bit
energy required at zero spectral efficiency and wideband slope.

We first considered the case in which the input, in addition to
the average power constraint, is subject to a fourth moment
constraint $E\{|x|^4\} \le \kappa P_\av^2$. We have shown that if
$\kappa > \left(1+\K\right)^2$, the wideband slope is negative.
Hence, the minimum bit energy is achieved at some nonzero spectral
efficiency, ${\sf{C}}^*$. In this case, we have identified a
forbidden region where one should not operate. In this region
where ${\sf{C}} < {\sf{C}}^*$, decreasing the spectral efficiency
further (i.e., increasing the bandwidth for fixed rate
transmission) increases the bit energy required for reliable
communications. Indeed, for the unknown Rayleigh fading channel,
the bit energy at zero spectral efficiency is infinite.

If $\kappa \leq \left(1+\K\right)^2$, the wideband slope is
positive and we have conjectured that for large enough Rician
factor $\K$, the minimum bit energy is achieved at zero spectral
efficiency. This bit energy can be achieved by BPSK signaling.

We have also analyzed the case where the input peakedness is
restricted by a peak power constraint. If the input
peak-to-average power ratio is limited, i.e., the input is subject
to $|x|^2 \stackrel{\text{a.s.}}{\le} \kappa P_\av$, the same
expressions for the bit energy at zero spectral efficiency and
wideband slope are found as in the fourth moment limited case.
Therefore the same conclusions as above are drawn. If the input
subject to a fixed peak limit, i.e., $|x|^2
\stackrel{\text{a.s.}}{\le} \nu$, we have obtained the minimum bit
energy and wideband slope. We have shown that if
$\frac{\gamma^2}{N_0}\nu \ge 1$, then the wideband slope is zero,
and hence achieving the minimum bit energy is very demanding in
bandwidth.

We have defined the OOBPSK and OOQPSK signaling schemes and
analyzed their low-power performance. We have shown that when the
input is subject to $E\{|x|^4\} \le \kappa P_\av^2$ or $|x|^2
\stackrel{\text{a.s.}}{\le} \kappa P_\av$, OOQPSK is second-order
optimal while OOBPSK is first-order optimal. Therefore, achieving
the optimal wideband slope, OOQPSK signaling turns out to be a
very efficient scheme in the low-power regime. In the case of the
fixed peak limit, we have seen that OOK signaling with on-level
fixed at the peak level achieves the minimum bit energy.

\appendix

\section{Proof of Proposition \ref{prop:Ricianwithpeaklevel}}
\label{app:proofofprop3}

Since the capacity curve is a concave function of the \tsnr in
this case, the minimum received bit energy is achieved at zero
spectral efficiency and can be obtained from $
\frac{E_b^r}{N_0}_{\min} =
\frac{(|m|^2+\gamma^2)\log_e2}{\dot{C}(0)}. $ $\dot{C}(0)$ is
easily found using the following formula \cite{Verdu_cost}
\begin{align}
\dot{C}(0) = N_0 \sup_{|x_0| \le \nu} \frac{D(f_{y|x=x_0}\|
f_{y|x=0})}{|x_0|^2} &= N_0 \sup_{|x_0| \le \nu} \frac{\frac{|m|^2
+ \gamma^2}{N_0} |x_0|^2 -
\log_e\left(\frac{\gamma^2}{N_0}|x_0|^2+1 \right)}{|x_0|^2} \\ &=
|m|^2 + \gamma^2 - N_0 \frac{\log_e\left(\frac{\gamma^2}{N_0}\nu
+1\right)}{\nu}. \label{eq:firstderivative}
\end{align}
Next we show that input signaling that satisfies
(\ref{eq:optcondition1}) and (\ref{eq:optcondition2}) is required
to achieve the minimum bit energy, and hence the optimal wideband
slope. We adopt an approach similar to that of \cite{Verdu} where
{\em{flash signaling}} is shown to be necessary to achieve the
minimum bit energy in the absence of the peak constraint
(\ref{eq:peakconstraintlimitedpeak}).
\\
\textbf{i}) We first prove that signaling that satisfies
(\ref{eq:optcondition1}) and (\ref{eq:optcondition2}) achieves the
first derivative of the capacity. Note that in general we have
\begin{align} \dot{I}(0) &= \lim_{\ssnr \to 0}
\frac{I(x_\ssnr;y)}{\ssnr} \\ &= \lim_{\ssnr \to 0} N_0
\frac{E\{D(f_{y|x = x_{\text{SNR}}}\| f_{y|x = 0
})\}}{E\{|x_\ssnr|^2\}} \\ &= |m|^2 + \gamma^2 - \lim_{\ssnr \to
0} N_0 \frac{E\left\{\log_e\left( \frac{\gamma^2}{N_0}|x_\ssnr|^2
+ 1 \right)\right\}}{E\{|x_\ssnr|^2\}}.
\label{eq:peaksignalfirstderiv}
\end{align}
Fix some $\epsilon \in (0,\nu)$. Then we can write
\begin{align}
E\left\{\log_e\left( \frac{\gamma^2}{N_0}|x_\ssnr|^2 + 1
\right)\right\} = &E\left\{\log_e\left(
\frac{\gamma^2}{N_0}|x_\ssnr|^2 + 1 \right)\,\,
1\left\{|x_\ssnr|^2
> \nu - \epsilon\right\}\right\} \nonumber \\ &+ E\left\{\log_e\left(
\frac{\gamma^2}{N_0}|x_\ssnr|^2 + 1 \right) \,\,
1\left\{|x_\ssnr|^2 \le \nu - \epsilon\right\}\right\}.
\label{eq:epsilonpartition}
\end{align}
Using $\log_e(1+x) \le x$, we have
\begin{align}
\lim_{\ssnr \to 0} \frac{E\left\{\log_e\left(
\frac{\gamma^2}{N_0}|x_\ssnr|^2 + 1 \right)\,\,1\left\{|x_\ssnr|^2
\le \nu - \epsilon\right\}\right\}}{E\{|x_\ssnr|^2\}} &\le
\frac{\gamma^2}{N_0} \lim_{\ssnr \to 0} \frac{E\left\{|x_\ssnr|^2
\,\,1\{|x_\ssnr|^2 \le \nu - \epsilon
\}\right\}}{E\{|x_\ssnr|^2\}} \nonumber \\ &= 0
\label{eq:upperzero}
\end{align}
where (\ref{eq:upperzero}) follows from (\ref{eq:optcondition2}).
Moreover using the fact that $\frac{\log_e(1+x)}{x}$ is
monotonically decreasing, $|x_\ssnr|^2 \stackrel{a.s.}\le \nu$,
and (\ref{eq:optcondition2}), we easily observe that
\begin{gather}
\frac{\log_e\left(\frac{\gamma^2}{N_0}\nu
 + 1\right)}{\nu} \le \lim_{\ssnr \to 0} \frac{E\left\{\log_e\left(
\frac{\gamma^2}{N_0}|x_\ssnr|^2 + 1 \right)\,\,1\left\{|x_\ssnr|^2
> \nu - \epsilon\right\}\right\}}{E\{|x_\ssnr|^2\}} \le \frac{\log_e\left(\frac{\gamma^2}{N_0}(\nu - \epsilon)
 + 1\right)}{\nu - \epsilon}. \label{eq:upperlower}
\end{gather}
From (\ref{eq:peaksignalfirstderiv}), (\ref{eq:epsilonpartition}),
(\ref{eq:upperzero}) and (\ref{eq:upperlower}), we have
\begin{gather}
|m|^2 + \gamma^2 - N_0 \frac{\log_e\left(\frac{\gamma^2}{N_0}(\nu
- \epsilon)+ 1\right)}{\nu - \epsilon} \le \dot{I}(0) \le |m|^2 +
\gamma^2 - N_0 \frac{\log_e\left(\frac{\gamma^2}{N_0}\nu+
1\right)}{\nu}.
\end{gather}
Since $\epsilon$ is arbitrary, we conclude that input signaling
satisfying (\ref{eq:optcondition1}) and (\ref{eq:optcondition2})
achieves $\dot{C}(0)$.
\\
\textbf{ii}) Now we will show that (\ref{eq:optcondition1}) and
(\ref{eq:optcondition2}) are necessary conditions to achieve
$\dot{C}(0)$. Note that (\ref{eq:optcondition1}) is dictated by
the input constraints (\ref{eq:constraint1}) and
(\ref{eq:peakconstraintlimitedpeak}). Using again the monotonicity
of $\frac{\log_e(1+x)}{x}$, we observe for arbitrary $\epsilon \in
(0,\nu)$ that
\begin{align}
E\left\{\log_e\left( \frac{\gamma^2}{N_0}|x_\ssnr|^2 + 1
\right)\,1\left\{|x_\ssnr|^2
> \nu - \epsilon\right\}\right\} &\ge \frac{\log_e\left(\frac{\gamma^2}{N_0}\nu +1
\right)}{\nu} E\left\{|x_\ssnr|^2 \,1\left\{|x_\ssnr|^2
> \nu - \epsilon\right\}\right\}, \label{eq:lowerboundonpartition1}
\\
E\left\{\log_e\left( \frac{\gamma^2}{N_0}|x_\ssnr|^2 + 1
\right)\,1\left\{|x_\ssnr|^2 \le \nu - \epsilon\right\}\right\}
&\ge \frac{\log_e\left(\frac{\gamma^2}{N_0}(\nu-\epsilon) +1
\right)}{\nu-\epsilon} E\left\{|x_\ssnr|^2 \,1\left\{|x_\ssnr|^2
\le \nu - \epsilon\right\}\right\}.
\label{eq:lowerboundonpartition2}
\end{align}
Combining (\ref{eq:peaksignalfirstderiv}),
(\ref{eq:epsilonpartition}), (\ref{eq:lowerboundonpartition1}) and
(\ref{eq:lowerboundonpartition2}), we have
\begin{align}
\dot{I}(0) \le \,\,&\,\,|m|^2 + \gamma^2 -
\frac{\log_e\left(\frac{\gamma^2}{N_0}\nu +1 \right)}{\nu}
\nonumber
\\
&- \left[\frac{\log_e\left(\frac{\gamma^2}{N_0}(\nu-\epsilon) +1
\right)}{\nu-\epsilon} - \frac{\log_e\left(\frac{\gamma^2}{N_0}\nu
+1 \right)}{\nu} \right] \lim_{\ssnr \to 0}
\frac{E\left\{|x_\ssnr|^2 \,1\left\{|x_\ssnr|^2 \le \nu -
\epsilon\right\}\right\}}{E\{|x_\ssnr|^2\}},
\label{eq:lowerboundonoverall}
\end{align}
from which we notice that the condition $\lim_{\ssnr \to 0}
\frac{E\left\{|x_{\text{SNR}}|^2 \,1\left\{|x_{\text{SNR}}|^2 \le
\nu - \epsilon\right\}\right\}}{E\{|x_{\text{SNR}}|^2\}} = 0$ for
all $0 < \epsilon <\nu$, which is equivalent to
(\ref{eq:optcondition2}), is required to achieve $\dot{C}(0)$.
\\ \\
\textbf{iii}) In this part, we obtain the optimal wideband slope
by evaluating $\ddot{C}(0)$. For the input $x_\ssnr$ that achieves
both $\dot{C}(0)$ and $\ddot{C}(0)$, we can write
\begin{gather}
\ddot{C}(0) = 2 \lim_{\ssnr \to 0} \frac{I(x_\ssnr;y) - \dot{C}(0)
\ssnr }{\ssnr^2} = -2 \lim_{\ssnr \to 0} \frac{D(f_y\|f_{y|x = 0
})}{\ssnr^2}. \label{eq:secondderivativewithdivergence}
\end{gather}
Furthermore by Proposition 1 of \cite{gursoy}, we can assume
without loss of optimality that $x_\ssnr$ has uniformly
distributed phase independent of the amplitude. With this
assumption, it can be easily verified that %\vspace{-.6cm}
\begin{gather}
D(f_y\|f_{y|x = 0 }) = D(f_R \| f_{R|r=0})
\end{gather}
where, as in \cite{gursoy}, $R = \frac{|y|^2}{N_0}$ and $r =
\frac{\gamma}{\sqrt{N_0}} \, x_{\text{SNR}}$, and therefore $f_R =
\int_{0}^{\infty} g(R,r) \, \ud F_r(r)$ with $g(R,r) =
\frac{1}{1+r^{2}} \exp \left( -\frac{R+{\sf{K}}r^{2}}{1+r^{2}}
\right) I_0 \left( \frac{2\sqrt{{\sf{K}}}r \sqrt{R}}{1 + r^{2}}
\right)$, and $f_{R|r = 0} = \exp(-R)$. Following the approach
employed in the proof of Theorem 16 in \cite{Verdu}, we write
%\vspace{-.2cm}
\begin{gather}
D(f_R \| f_{R|r=0}) = E\{(1 + \ssnr (W+V)) \log_e (1 + \ssnr
(W+V))\}
\end{gather}
%\vspace{-.3cm}
where %\vspace{-.3cm}
\begin{align}
V = \frac{P(r^2 > \eta - \epsilon)}{\ssnr}
\left(\frac{\tilde{f}_R}{f_{R|r=0}} -1 \right) \quad \text{and}
\quad W = \frac{1 - P(r^2 > \eta - \epsilon)}{\ssnr}
\left(\frac{\hat{f}_R}{f_{R|r=0}} -1 \right).
\end{align}
In the above formulation $\tilde{f}_R$ and $\hat{f}_R$ are the
distributions of $R$ conditioned on $r^2 > \eta - \epsilon$ and
$r^2 \le \eta - \epsilon$, respectively, for some fixed $\e \in
(0,\eta)$. Using the facts that $(x+1)\log_e(1+x) = x +
\frac{1}{2}x^2 + o(x^2)$, and $(V+W)$ has zero mean, and converges
to a nonzero random variable for vanishing $\tsnr\!\!$ when
(\ref{eq:optcondition1}) and (\ref{eq:optcondition2}) are
satisfied, we have
\begin{align}
\lim_{\ssnr \to 0} \frac{D(f_R \| f_{R|r=0})}{\ssnr^2} &=
\lim_{\ssnr \to 0} \frac{E\{(1 + \ssnr (W+V)) \log_e (1 + \ssnr
(W+V))\}}{\ssnr^2}
\\
&= \frac{1}{2} \lim_{\ssnr \to 0} E\{(W+V)^2\}.
\label{eq:asymptoticlowerboundondivergence}
\end{align}
Noting that (\ref{eq:optcondition1}) and (\ref{eq:optcondition2})
are necessary to achieve the minimum bit energy, and hence the
optimal wideband slope, we will first consider $E\{V^2\}$.
\begin{align}
\lim_{\ssnr \to 0} E\{V^2\} &= \lim_{\ssnr \to 0} \frac{P^2(r^2 >
\eta - \epsilon)}{\ssnr^2}
E\left\{\frac{\tilde{f}_R^2}{f^2_{R|r=0}}-1\right\}
\\
&\ge \frac{N_0^2}{\nu^2} \lim_{\ssnr \to
0}E\left\{\frac{\tilde{f}_R^2}{f^2_{R|r=0}}-1\right\}
\label{eq:lowerboundonprobability}
\\
&= \frac{N_0^2}{\nu^2} \lim_{\ssnr \to 0} \left(\int e^R
\tilde{f}^2_R \, \ud R - 1 \right)
\\
&\ge \frac{N_0^2}{\nu^2} \left( \int_{\Omega_1} e^R
g^2(R,\sqrt{\eta}) \, \ud R + \int_{\mathbb{R}^+ \setminus
\Omega_1} e^R g^2(R,\sqrt{\eta - \epsilon}) \, \ud R - 1 \right),
\label{eq:lowerboundonintegral}
\end{align}
where $\Omega_1 = \{R: g(R,\sqrt{\eta}) \le
g(R,\sqrt{\eta-\epsilon})\}$, and $\mathbb{R}^+ = [0,\infty)$.
(\ref{eq:lowerboundonprobability}) follows by assuming
(\ref{eq:optcondition2}) and noting that
\begin{align}
1 = \lim_{\ssnr \to 0} \frac{E\{|x_\ssnr|^2\} \,\, 1\{|x_\ssnr|^2
> \nu - \epsilon \}}{E\{|x_\ssnr|^2\}} &= \lim_{\ssnr \to 0}
\frac{P(|x_\ssnr|^2 > \nu - \epsilon)E\{|x_\ssnr|^2\}\,\, |
\,\,|x_\ssnr|^2
> \nu - \e \}}{E\{|x_\ssnr|^2\}} \nonumber
\\
&\le \nu \lim_{\ssnr \to 0} \frac{P(|x_\ssnr|^2 > \nu -
\epsilon)}{E\{|x_\ssnr|^2\}}
\\
&= \frac{\nu}{N_0} \lim_{\ssnr \to 0} \frac{P(|x_\ssnr|^2 > \nu -
\epsilon)}{\ssnr}.
\end{align}
(\ref{eq:lowerboundonintegral}) follows by noting that
\begin{align}
\tilde{f}_r &= \int g(R,r) \, \ud \tilde{F}_r(r)
\\
&\ge \min\{g(R,\sqrt{\eta}), g(R,\sqrt{\eta - \e})\}
\label{eq:lowerboundong}
\end{align}
where $\tilde{F}_r$ is the distribution of $r$ conditioned on $r^2
> \eta - \e$.
(\ref{eq:lowerboundong}) follows from the fact that $g(R,r)$ is
either a monotonically decreasing or a first monotonically
increasing and then decreasing function of $r$.

Similar analysis leads to $\lim_{\ssnr \to 0} E\{WV\} = 0$.
Therefore from (\ref{eq:secondderivativewithdivergence}),
(\ref{eq:asymptoticlowerboundondivergence}), and
(\ref{eq:lowerboundonintegral}), we have
\begin{align}
\ddot{C}(0) &\le - \frac{N_0^2}{\nu^2} \left( \int_{\Omega_1} e^R
g^2(R,\sqrt{\eta}) \, \ud R + \int_{\mathbb{R} \setminus \Omega_1}
e^R g^2(R,\sqrt{\eta - \epsilon}) \, \ud R - 1 \right).
\end{align}
As $\e \in (0,\eta)$ is arbitrary, we have
\begin{align}
\ddot{C}(0) \le - \frac{N_0^2}{\nu^2} \left( \int_0^{\infty} e^R
g^2(R,\sqrt{\eta}) \, \ud R  - 1\right).
\label{eq:upperboundonsecondderivative}
\end{align}
It can be easily shown that the two-mass-point input amplitude
distribution %\vspace{-.5cm}
\begin{gather}
F(|x|) = \left( 1 - \frac{P_\av}{\nu} \right) u(|x|) +
\frac{P_\av}{\nu}\,u(|x| - \sqrt{v}), \label{eq:optimalinput}
\end{gather}
achieves both $\dot{C}(0)$ and the upper bound in
(\ref{eq:upperboundonsecondderivative}). Therefore we conclude
that (\ref{eq:upperboundonsecondderivative}) is indeed satisfied
with equality, and therefore we have
\begin{gather}
\ddot{C}(0) = \left\{
\begin{array}{ll}
-\frac{N_0^2}{\nu^2} \left(
\frac{1}{1-\frac{\gamma^4\nu^2}{N_0^2}} \exp\left( \frac{2
\frac{|m|^2\gamma^2\nu^2}{N_0^2}}{1 -
\frac{\gamma^4\nu^2}{N_0^2}}\right) I_0\left(\frac{2
\frac{|m|^2\nu}{N_0}}{1-\frac{\gamma^4\nu^2}{N_0^2}}\right) - 1
\right) & \nu < \frac{N_0}{\gamma^2} \\ -\infty & \nu \ge
\frac{N_0}{\gamma^2}
\end{array} \right.
\label{eq:secondderivativepeaklimit}
\end{gather}
where (\ref{eq:secondderivativepeaklimit}) is obtained by
evaluating a closed form expression for the integral in
(\ref{eq:upperboundonsecondderivative}). The wideband slope is
obtained by inserting (\ref{eq:firstderivative}) and
(\ref{eq:secondderivativepeaklimit}) into (\ref{eq:widebanddef}).
\hfill $\square$

\begin{spacing}{0.8}

\end{spacing}
\newpage


\end{spacing}
\end{document}